\documentclass[twocolumn, showpacs, nofootinbib, aps, prd]{revtex4}

\usepackage{graphicx}
\usepackage{dcolumn}
\usepackage{color}
\usepackage{amsfonts}
\usepackage[colorlinks=true,linkcolor=blue,citecolor=blue, urlcolor=blue]{hyperref}

\begin{document}

\title{Heavy Pseudoscalar Twist-3 Distribution Amplitudes within QCD Theory in Background Fields}

\author{Tao Zhong$^{1}$}
\email{zhongtao@htu.edu.cn}
\author{Xing-Gang Wu$^{2}$}
\email{wuxg@cqu.edu.cn}
\author{Tao Huang$^{3}$}
\email{huangtao@ihep.ac.cn}
\author{Hai-Bing Fu$^{4}$}
\email{fuhb@cqu.edu.cn}

\address{$^1$ Department of Physics, Henan Normal University, Xinxiang 453007, P.R. China\\
$^2$ Department of Physics, Chongqing University, Chongqing 401331, P.R. China \\
$^3$ Institute of High Energy Physics and Theoretical Physics Center for Science Facilities, Chinese Academy of Sciences, Beijing 100049, P.R. China \\
$^4$ School of Science, Guizhou Minzu University, Guiyang 550025, P. R. China}

\date{\today}

\begin{abstract}

In this paper, we study the properties of the twist-3 distribution amplitude (DA) of the heavy pseudo-scalars such as $\eta_c$, $B_c$ and $\eta_b$. New sum rules for the twist-3 DA moments $\left<\xi^n_P\right>_{\rm HP}$ and $\left<\xi^n_\sigma\right>_{\rm HP}$ up to sixth orders and up to dimension-six condensates are deduced under the framework of the background field theory. Based on the sum rules for the twist-3 DA moments, we construct a new model for the two twist-3 DAs of the heavy pseudo-scalar with the help of the Brodsky-Huang-Lepage prescription. Furthermore, we apply them to the $B_c\to\eta_c$ transition form factor ($f^{B_c\to\eta_c}_+(q^2)$) within the light-cone sum rules approach, and the results are comparable with other approaches. It has been found that the twist-3 DAs $\phi^P_{3;\eta_c}$ and $\phi^\sigma_{3;\eta_c}$ are important for a reliable prediction of $f^{B_c\to\eta_c}_+(q^2)$. For example, at the maximum recoil region, we have $f^{B_c\to\eta_c}_+(0) = 0.674 \pm 0.066$, in which those two twist-3 terms provide $\sim33\%$ and $\sim22\%$ contributions. Also we calculate the branching ratio of the semi-leptonic decay $B_c \to\eta_c l\nu$ as $Br(B_c \to\eta_c l\nu) = \left( 9.31^{+2.27}_{-2.01} \right) \times 10^{-3}$.

\end{abstract}

\pacs{11.55.Hx, 14.40.-n}

\maketitle

\section{introduction}

The heavy pseudo-scalar (HP), such as $B_c$, $\eta_c$ or $\eta_b$, is a ground-state meson constituted by a heavy quark and an anti-quark. For exclusive process involving the HP, its distribution amplitude (DA) is usually key component for predicting the decay widths or the production cross-sections of those processes. Thus a more precise HP DA shall lead to a more precise predictions.

Due to non-relativistic nature of the heavy constituent quark/antiquark, the leading-twist DA (or the twist-2 DA) of the HP can be roughly treated as a $\delta$-function~\cite{BC_DA}, i.e. in the leading order of the expansion over the relative velocities, the quark and the antiquark in the HP DA simply share the momentum of the meson according to their masses. For examples, the asymptotic DA of the charmonium and the bottomonium are $\phi^{\rm asy}_{2;\rm HP}(x) \sim \delta(x-1/2)$. By taking the relativistic effect, the heavy-quark mass effect and/or the three-particle effect into account, such a simple asymptotic DA shall be broadened to a certain degree, cf. Refs.\cite{BC_DA, Kawamura:2001jm, Huang:2005kk}. Several models for a broader and more realistic HP twist-2 DA have been suggested in the literature, such as the Bondar-Chernyak (BC) model~\cite{ETAC_T2DA_BC}, the Bodwin-Kang-Lee (BKL) model~\cite{ETAC_T2DA_BKL}, the Ma-Si (MS) model~\cite{ETAC_T2DA_MS}, the Braguta-Likhoded-Luchinsky (BLL) model~\cite{ETAC_T2DA_BLL}, the model with the Brodsky-Huang-Lepage (BHL) prescription~\cite{ETAC_T2DA_BHL1, ETAC_T2DA_BHL2}, and etc. A comparison of various $\eta_c$ twist-2 DA has been done in Ref.\cite{Sun:2009zk}, in which it has also pointed out that by using a proper $\eta_c$ twist-2 DA, one may resolve the disagreement between the experimental observations and the NRQCD prediction on the production cross-section of $e^+ e^- \to J/\Psi+ \eta_c$.

The background field theory (BFT)~\cite{BFT1, BFT2, BFT3} provides a systematic approach for achieving the goal of SVZ sum rules~\cite{SVZ} and also provides a physical picture for the vacuum condensates. In Ref.~\cite{HP_T2DA_zhong}, by using the SVZ sum rules within the framework of BFT, we have made a detailed study on the properties of the HP twist-2 DA and have constructed a new model for the twist-2 DA. According to our knowledge on the pionic cases, in addition to the twist-2 DA, the high-twist DA may also provide sizable contributions to the HP-involved processes, even though they are generally power suppressed. Taking the $B\to\pi$ transition form factor (TFF) as an example, a large contribution from the pion twist-3 DAs $\phi^P_{3;{\pi}}$ and $\phi^\sigma_{3;{\pi}}$ in comparison to its twist-2 DA has been observed in both the intermediate and the large $Q^2$-region~\cite{ht1, ht2, ht3, ht4}. It is thus interesting to have a reliable way to estimate the properties of the HP twsit-3 DA, which may also have sizable contributions to high-energy processes but are less known so far.

In the present paper, as a step forward, we shall study the HP twist-3 DAs $\phi^P_{3;{\rm HP}}$ and $\phi^\sigma_{3;{\rm HP}}$ by using the same approach of Ref.\cite{HP_T2DA_zhong}. We shall first estimate the moments of $\phi^P_{3;{\rm HP}}$ and $\phi^\sigma_{3;{\rm HP}}$ up to dimension-six by using the SVZ sum rules within the framework of BFT. Then, we shall construct a model for $\phi^P_{3;{\rm HP}}$ and $\phi^\sigma_{3;{\rm HP}}$ with the help of the BHL-prescription. In this way the HP twist-3 DAs with a better end-point behavior can be achieved. Finally, as an application of the constructed HP twist-3 DAs, we shall apply them to the $B\to \eta_c$ TFF $f^{B_c\to\eta_c}_+(q^2)$.

The remaining parts of the paper are organized as follows. In Sec.II, we construct a light-cone harmonic oscillator model for the HP twist-3 DA with the help of the BHL-prescription. The DA moments are calculated with SVZ sum rules within the framework of BFT. The corresponding numerical results are presented in Sec.III. As an application of the suggested HP twist-3 DA model, we give the light-cone sum rules (LCSR) for $f^{B_c\to\eta_c}_+(q^2)$ by using the conventional correlator, and a comparison of $f^{B_c\to\eta_c}_+(q^2)$ with other approaches are also given in Sec.IV. Finally,Sec.V is reserved for a summary.

\section{calculation technique}

\subsection{Model for the HP Twist-3 DA}

Based on the BHL-prescription~\cite{BHL}, a light-cone harmonic oscillator model for the HP twist-2 wave function (WF) has been built in Ref.~\cite{HP_T2DA_zhong}~\footnote{For the detail technique, we refer the reader to Ref.~\cite{BFTSR}, where the quark propagator is given with full mass dependence up to dimension-six terms within the framework of BFT.}. After integrating out the transverse momentum ${\textbf k}_\perp$ component of the WF, the corresponding DA is achieved. Similarly, the HP twist-3 DA $\phi^P_{3;\rm HP}$ can be constructed as
\begin{eqnarray}
\phi^P_{3;\rm HP} (x,\mu_0) &=& \frac{\sqrt{6} A_{\rm HP}^P (\beta_{\rm HP}^{P})^2}{\pi^2 f_{\rm HP}} x(1-x) \varphi_{\rm HP}^P(x) \nonumber\\
&\times& \exp\left[ -\frac{\hat{m}_1^2(1-x) + \hat{m}_2^2x}{8(\beta_{\rm HP}^{P})^2 x(1-x)} \right] \nonumber\\
&\times& \left( 1 - \exp\left[ -\frac{\mu_0^2}{8(\beta_{\rm HP}^{P})^2 x(1-x)} \right] \right), 
\label{DA_model_P}
\end{eqnarray}
where $f_{\rm HP}$ stands for the HP decay constant, $\hat{m}_{1,2}$ are constituent quark masses of the HP, $\mu_0$ is the factorization scale, and $\textrm{Erf}(x) = \frac{2}{\sqrt{\pi}} \int^x_0 e^{-t^2} dt$ is the error function. The constituent quark masses $\hat{m}_1= \hat{m}_b$ and $\hat{m}_2 = \hat{m}_c$ for the case of $B_c$-meson and $\hat{m}_1 = \hat{m}_2 = \hat{m}_c (\hat{m}_b)$ for the case of $\eta_c (\eta_b)$-meson. We take $\hat{m}_c = 1.8 {\rm GeV}$ and $\hat{m}_b = 4.7 {\rm GeV}$ to do our numerical calculation. $A^P_{\rm HP}$ is the normalization constant. The harmonious parameter ${\beta^P_{\rm HP}}$ dominantly determines the transverse distribution of the DA. The function $\varphi^P_{\rm HP}(x)$ dominates the DA longitudinal distribution, which is expressed by a Gegenbauer polynomial. Keeping its first several terms, we have
\begin{eqnarray}
\varphi^P_{\rm HP}(x) = 1 + \sum^6_{n=1} B^{{\rm HP},P}_n \times C^{1/2}_n(2x-1). \label{varphi_p}
\end{eqnarray}
Due to the same mass for the heavy constituent quark and anti-quark, the twist-3 DA $\phi^P_{3;\eta_c}$ and $\phi^P_{3;\eta_b}$ of $\eta_c$ and $\eta_b$ mesons are symmetric under the transformation of $x \leftrightarrow (1-x)$, thus we have $B^{\eta_c,P}_{2m-1} = B^{\eta_b,P}_{2m-1} = 0$ for $(m\geq1)$.

The input parameters $A^P_{\rm HP}$, $B_n^{{\rm HP},P}$ and $\beta^P_{\rm HP}$ are determined by the following constraints:
\begin{itemize}
\item The normalization condition of $\phi^P_{3;\rm HP}$,
\begin{eqnarray}
\int^1_0 dx \phi^P_{3;\rm HP}(x,\mu_0) = 1.
\label{norma_condi}
\end{eqnarray}

\item The average value of the squared HP transverse momentum $\left<\textbf{k}_\bot^2\right>_{\rm HP}$, which is defined as
\begin{eqnarray}
\left<\textbf{k}_\perp^2\right>_{\rm HP} &=& \frac{(A_{\rm HP}^{P})^2 (\beta_{\rm HP}^{P})^4}{\pi^2 P_{\rm HP}} \int^1_0 dx x^2(1-x)^2 \nonumber\\
&\times& (\varphi_{\rm HP}^{P}(x))^2 \exp \left[ -\frac{\hat{m}_1^2(1-x) + \hat{m}_2^2x}{4(\beta_{\rm HP}^{P})^2 x(1-x)} \right],
\label{avera_value_k}
\end{eqnarray}
where $P_{\rm HP}$ stands for the probability of finding the valence quark state $\left|Q_1\bar{Q}_2\right>$ in the HP Fock-state expansion, and we take, $P_{\eta_c} \simeq 0.8$ and $P_{B_c}\sim P_{\eta_b} \simeq 1$~\cite{HP_T2DA_zhong}. At present, there is no definite value for $\left<\textbf{k}_\bot^2\right>_{\rm HP}$, and we predict its value via a comparison of the cases of the pion and the $B/D$ mesons that are better known. The average values of the squared transverse momentum of $D$ and $B$ mesons are sensitive to their decay constants $f_D$ and $f_B$~\cite{AK2_BD}, $\left<\textbf{k}^2_\bot \right>^{1/2}_{D(B)} \propto f_{D(B)}$. It is reasonable to assume that this is also satisfied by the HP meson, $\left<\textbf{k}_\bot^2\right>_{\rm HP}^{1/2} \propto f_{\rm HP}$. Moreover, we connect the HP value $\left<\textbf{k}_\bot^2\right>_{\rm HP}^{1/2}$ to the case of pion (the light pseudoscalar) as, $\left<\textbf{k}_\bot^2\right>_{\rm HP}^{1/2} / f_{\rm HP} \simeq \left<\textbf{k}_\bot^2\right>_\pi^{1/2} / f_\pi$, where $f_\pi = (130.41 \pm 0.03 \pm 0.20)$ MeV~\cite{PDG} is the pion decay constant. By further taking $\left<\textbf{k}_\bot^2\right>_\pi^{1/2} \simeq 350 {\rm MeV}$~\cite{AK2_PI, AK2_BD}, we finally get $\left<\textbf{k}_\bot^2\right>_{\eta_c}^{1/2} \simeq 1.216$ GeV, $\left<\textbf{k}_\bot^2\right>_{B_c}^{1/2} \simeq 1.337$ GeV and $\left<\textbf{k}_\bot^2\right>_{\eta_b}^{1/2} \simeq 2.177$ GeV. It is noted that the shape of $\phi^P_{3;\rm HP}$ (or $\phi^\sigma_{3;\rm HP}$) is dominated by their moments and is insensitive to the choice of $\left<\textbf{k}_\bot^2\right>_{\rm HP}^{1/2}$.

\item The twist-3 DA moments $\left<\xi^n_P\right>_{\rm HP}$ are defined as
\begin{eqnarray}
\left<\xi^n_P\right>_{\rm HP} |_{\mu_0} = \int^1_0 du (2u-1)^n \phi_{3;\rm HP}^P(u,\mu_0), \label{mom_t3ps}
\end{eqnarray}
which shall be calculated by using the SVZ sum rules within the framework of BFT in the next subsection.
\end{itemize}

The model of $\phi^\sigma_{3;\rm HP}$ can be constructed via the same way, i.e., by replacing the upper index `$P$' with `$\sigma$' in Eqs.(\ref{DA_model_P}, \ref{norma_condi}, \ref{avera_value_k} and \ref{mom_t3ps}), and taking the expansion
\begin{eqnarray}
\varphi^\sigma_{\rm HP}(x) = 1 + \sum^6_{n=1} B^{{\rm HP},\sigma}_n \times C^{3/2}_n(2x-1), \label{varphi_sig}
\end{eqnarray}
we can obtain the model for the HP twist-3 DA $\phi^\sigma_{3;\rm HP}$.

In the above equations, all the parameters are for the initial scale $\mu_0$, the parameters at any other scale can be obtained via the conventional evolution equation~\cite{HP_T2DA_zhong, EE}.

\subsection{Sum Rules for the Twist-3 DAs $\phi^P_{3;\rm HP}$ and $\phi^\sigma_{3;\rm HP}$}

To derive the sum rules for the moments of the twist-3 DAs $\phi^P_{3;\rm HP}$ and $\phi^\sigma_{3;\rm HP}$, we adopt the following correlators
\begin{eqnarray}
\Pi^{\rm PS}_{\rm HP}(q) &=& (z\cdot q)^{n} I^{\rm PS}_{\rm HP}(q^2) \nonumber\\
&=& i \int d^4x e^{iq\cdot x} \left<0\left| T \left\{ J^{\rm PS}_n(x) J^{{\rm PS}\dag}_0(0) \right\} \right|0\right>  \label{cor_xinps}
\end{eqnarray}
and
\begin{eqnarray}
\Pi^{\rm PT}_{\rm HP}(q) &=& - i (q_\mu z_\nu - q_\nu z_\mu)(z\cdot q)^{n} I^{\rm PT}_{\rm HP}(q)(q^2) \nonumber\\
&=& i \int d^4x e^{iq\cdot x} \left<0\left| T \left\{ J^{\rm PT}_n(x) J^{{\rm PS}\dag}_0(0) \right\} \right|0\right> \label{cor_xinpt}
\end{eqnarray}
for $\left<\xi^n_P\right>_{\rm HP}$ and $\left<\xi^n_\sigma\right>_{\rm HP}$, respectively. Here $z^2 = 0$, $J^{\rm PS}_n(x)$ and $J^{\rm PT}_n(x)$ stand for the pseudo-scalar and the tensor currents
\begin{eqnarray}
J^{\rm PS}_n(x) &=& \bar{Q}_1(x) \gamma_5 (iz\cdot \tensor{D})^n Q_2(x), \label{psjx}\\
J^{\rm PT}_n(x) &=& \bar{Q}_1(x) \sigma_{\mu\nu} \gamma_5 (iz\cdot \tensor{D})^{n+1} Q_2(x) \label{ptjx}
\end{eqnarray}
with $Q_{1}=b$ and $Q_{2}=c$ for $B_c$, $Q_{1}=Q_{2}=c$ ($Q_{1}=Q_{2}=b$) for $\eta_{c}$ ($\eta_{b}$), and $\sigma_{\mu\nu} = \frac{i}{2} (\gamma_\mu \gamma_\nu - \gamma_\nu \gamma_\mu)$.

Two correlators (\ref{cor_xinps}, \ref{cor_xinpt}) can be treated under the standard SVZ sum rules. On the one hand, in the physical region, one can insert a completed set of intermediate hadronic states in the correlators (\ref{cor_xinps}, \ref{cor_xinpt}). The hadronic transition matrix elements can be written as
\begin{eqnarray} &&
\left<0\left| \bar{Q}_1(0) \gamma_5 (iz\cdot \tensor{D})^n Q_2(0) \right|{\rm HP}(q)\right>
\nonumber\\ && \quad\quad\quad\quad\quad\quad
= -i \mu_{\rm HP} f_{\rm HP} (z\cdot q)^{n} \left<\xi^n_P\right>_{\rm HP}, \label{def_mom_t3ps}\\ &&
\left<0\left| \bar{Q}_1(0) \sigma_{\mu\nu} \gamma_5 (iz\cdot \tensor{D})^{n+1} Q_2(0) \right|{\rm HP}(q)\right>
\nonumber\\ && \quad\quad\quad\quad\quad\quad
= - \frac{n+1}{3} \mu_{\rm HP} f_{\rm HP} \left[ 1 - \frac{(m_1 + m_2)^2}{m_{\rm HP}^2} \right]
\nonumber\\ && \quad\quad\quad\quad\quad\quad
\times (q_\mu z_\nu - q_\nu z_\mu) (z\cdot q)^{n} \left<\xi^n_\sigma\right>_{\rm HP}. \label{def_mom_t3pt}
\end{eqnarray}
Furthermore, the hadronic spectrum representations can be written as
\begin{eqnarray}
{\rm Im} I^{\rm PS}_{\rm HP,had}(s) &=& \pi \delta (s - m_{\rm HP}^2) \mu_{\rm HP}^2 f^2_{\rm HP} \left<\xi^n_P\right>_{\rm HP} \nonumber\\
&+& \pi \rho^{\rm had}_P(s) \theta(s - s_{\rm HP}^P), \label{had_p}\\
{\rm Im} I^{\rm PT}_{\rm HP,had}(s) &=& \pi \delta (s - m_{\rm HP}^2) \frac{n+1}{3} \mu_{\rm HP}^2 f^2_{\rm HP} \nonumber\\
&\times& \left[ 1 - \frac{(m_1 + m_2)^2}{m_{\rm HP}^2} \right] \left<\xi^n_\sigma\right>_{\rm HP} \nonumber\\
&+& \pi \rho^{\rm had}_\sigma(s) \theta(s - s_{\rm HP}^\sigma),
\label{had_sig}
\end{eqnarray}
where $m_{\rm HP}$ is the HP mass, $\mu_{\rm HP} = \frac{m_{\rm HP}^2}{m_1 + m_2}$ with the mass $m_{1(2)}$ of the heavy quark $Q_{1(2)}$. $\theta$ stands for the usual step-function and $s_{\rm HP}^P$ and $s_{\rm HP}^\sigma$ indicate the continue threshold parameters. $\rho^{\rm had}_P (s)$ and $\rho^{\rm had}_\sigma (s)$ are hadronic spectrum densities, which can be approximately obtained with the quark-hadron duality~\cite{SVZ}.

On the other hand, the correlators (\ref{cor_xinps}, \ref{cor_xinpt}) can be treated by using the operator product expansion (OPE) in the deep Euclidean region. Detailed calculation technique for the OPE under the framework BFT can be found in Ref.\cite{HP_T2DA_zhong}. For shortness, we shall not present them here, the interesting reader may turn to Ref.\cite{HP_T2DA_zhong} for detail.

By further using the dispersion relation
\begin{eqnarray}
I_{\rm qcd}(q^2) = \frac{1}{\pi} \int^\infty_{t_{\rm min}} ds \frac{{\rm Im} I_{\rm had}(s)}{s - q^2} + {\rm subtractions},
\label{dis_rel}
\end{eqnarray}
where $t_{\rm min} = (m_1 + m_2)^2$, and applying the Borel transformation, we finally obtain the required sum rules for the moments $\left<\xi^n_P\right>_{\rm HP}$ and $\left<\xi^n_\sigma\right>_{\rm HP}$,
\begin{eqnarray} &&
\frac{\mu_{\rm HP}^2 f_{\rm HP}^2 \left<\xi^n_P\right>_{\rm HP}}{M^2 \exp \left[ m_{\rm HP}^2/M^2 \right]} \nonumber\\ &&
= \frac{1}{\pi} \frac{1}{M^2} \int^{s_0^P}_{t_{min}} ds e^{-s/M^2} {\rm Im} I^{\rm PS}_{\rm HP,pert}(s) + I^{\rm PS}_{{\rm HP},\left<G^2\right>}(M^2) \nonumber\\ &&
+ I^{\rm PS}_{{\rm HP},\left<G^3\right>}(M^2), \label{sr_p}\\ &&
\frac{(n+1)\mu_{\rm HP}^2 f_{\rm HP}^2 \left<\xi^n_\sigma\right>_{\rm HP}}{3M^2 \exp \left[ m_{\rm HP}^2/M^2 \right]} \times \left[ 1 - \frac{(m_1 + m_2)^2}{m_{\rm HP}^2} \right] \nonumber\\ &&
= \frac{1}{\pi} \frac{1}{M^2} \int^{s_0^\sigma}_{t_{min}} ds e^{-s/M^2} {\rm Im} I^{\rm PT}_{\rm HP,pert}(s) + I^{\rm PT}_{{\rm HP},\left<G^2\right>}(M^2) \nonumber\\ &&
+ I^{\rm PT}_{{\rm HP},\left<G^3\right>}(M^2), \label{sr_sig}
\end{eqnarray}
where $M$ stands for the Borel parameter. The functions ${\rm Im} I^{\rm PS,PT}_{\rm HP,pert}(s)$, $I^{\rm PS,PT}_{{\rm HP},\left<G^2\right>}(M^2)$ and $I^{\rm PS,PT}_{{\rm HP},\left<G^3\right>}(M^2)$ stand for imaginary part of the perturbative terms, the contribution proportional to double-gluon condensate $\left<G^2\right>$ and the contribution proportional to triple-gluon condensate $\left<g^3_sfG^3\right>$, respectively. For convenience, we put their expressions in the Appendix.

\section{The HP twist-3 DAs $\phi^P_{3;\rm HP}$ and $\phi^\sigma_{3;\rm HP}$}

To do the numerical calculation, we adopt the following Particle Data Group values for the input parameters~\cite{PDG}: $m_{\eta_c} = (2.9837 \pm 0.0007) \textrm{GeV}$, $m_{B_c} = (6.2745 \pm 0.0018) \textrm{GeV}$, $m_{\eta_b} = (9.3980 \pm 0.0032) \textrm{GeV}$; the current quark masses under the $\overline{\rm MS}$-scheme are $\bar{m}_c (\bar{m}_c) = (1.275 \pm 0.025) \textrm{GeV}$ and $\bar{m}_b (\bar{m}_b) = (4.18 \pm 0.03) \textrm{GeV}$. The one-loop $\alpha_s$-running is adopted, whose running behavior is fixed by $\alpha_s(m_Z) = 0.1185 \pm 0.0006$ with $m_Z = (91.1876 \pm 0.0021) \textrm{GeV}$~\cite{PDG}. And we obtain $\Lambda_{\rm QCD} \simeq 270\textrm{MeV}$, $257 \textrm{MeV}$ and $204 \textrm{MeV}$ for the flavor number $n_f = 3$, $4$ and $5$, respectively. The scale-independence gluon condensates are taken as $\left<\alpha_s G^2\right> = (0.038 \pm 0.011) \textrm{GeV}^4$~\cite{SVZ} and $\left<g_s^3 f G^3\right> = (0.013 \pm 0.007) \textrm{GeV}^6$~\cite{BFTSR}. As suggested by Braguta et al.~\cite{ETAC_T2DA_BLL}, the continuum threshold parameters $s_0^P$ and $s_0^\sigma$ are taken to be infinity, and the ratio $\left<\xi^n_P\right>_{\rm HP}/\left<\xi^0_P\right>_{\rm HP}$ and $\left<\xi^n_\sigma\right>_{\rm HP}/\left<\xi^0_\sigma\right>_{\rm HP}$ are adopted to derive the $n_{\rm th}$-moment $\left<\xi^n_P\right>_{\rm HP}$ and $\left<\xi^n_\sigma\right>_{\rm HP}$. As for the Borel windows of the sum rules (\ref{sr_p}, \ref{sr_sig}), we take $M^2 \in [1,2] ({\rm GeV}^2)$ for $\left<\xi^n_{P}\right>_{\eta_c}$ and $\left<\xi^n_{\sigma}\right>_{\eta_c}$, $M^2 \in [15,20] ({\rm GeV}^2)$ for $\left<\xi^n_{P}\right>_{B_c}$, $\left<\xi^n_{\sigma}\right>_{B_c}$, $\left<\xi^n_{P}\right>_{\eta_b}$ and $\left<\xi^n_{\sigma}\right>_{\eta_b}$.

\subsection{Properties of the HP twist-3 DA moments}

\begin{figure*}[tb]
\centering
\includegraphics[width=0.32\textwidth]{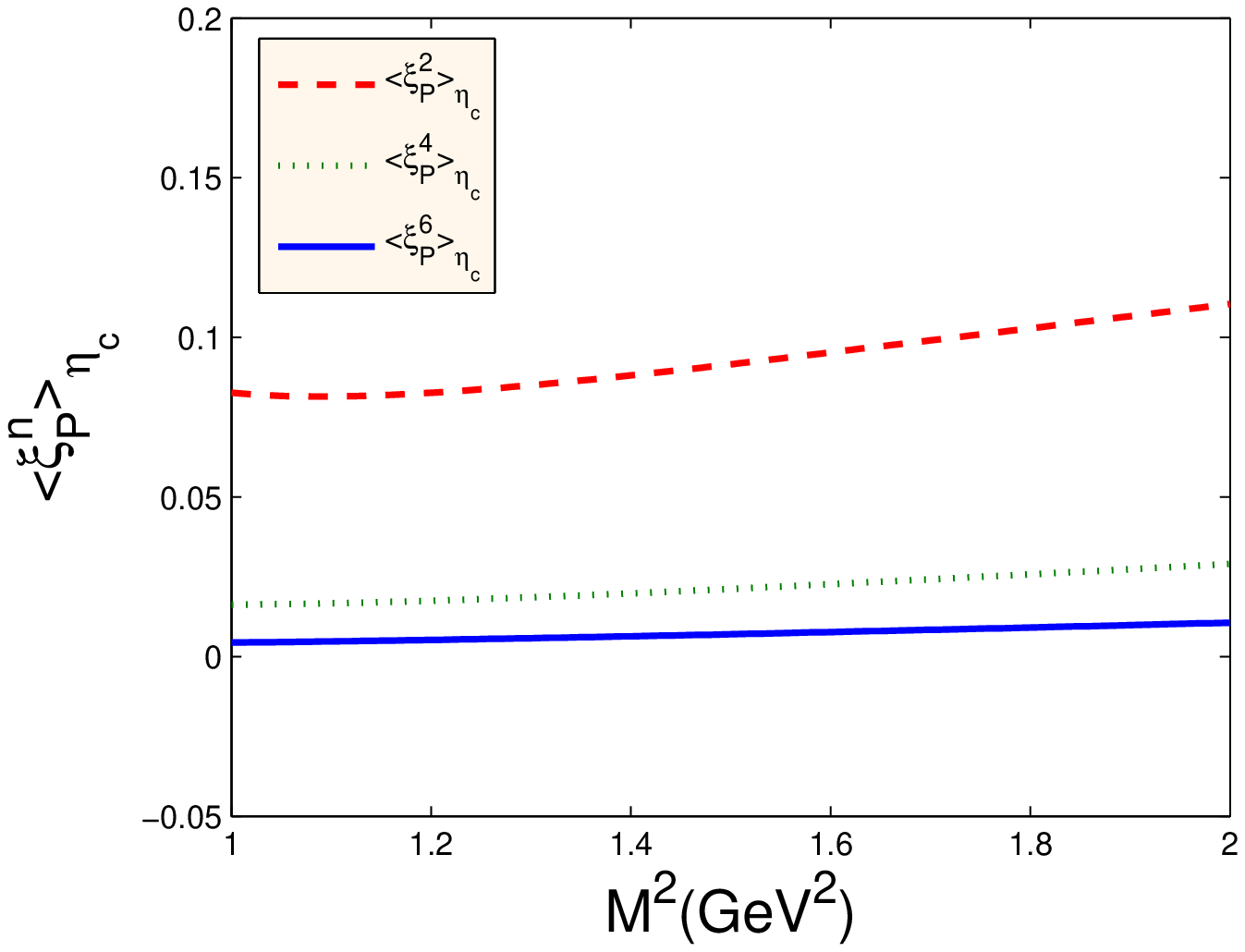}
\includegraphics[width=0.32\textwidth]{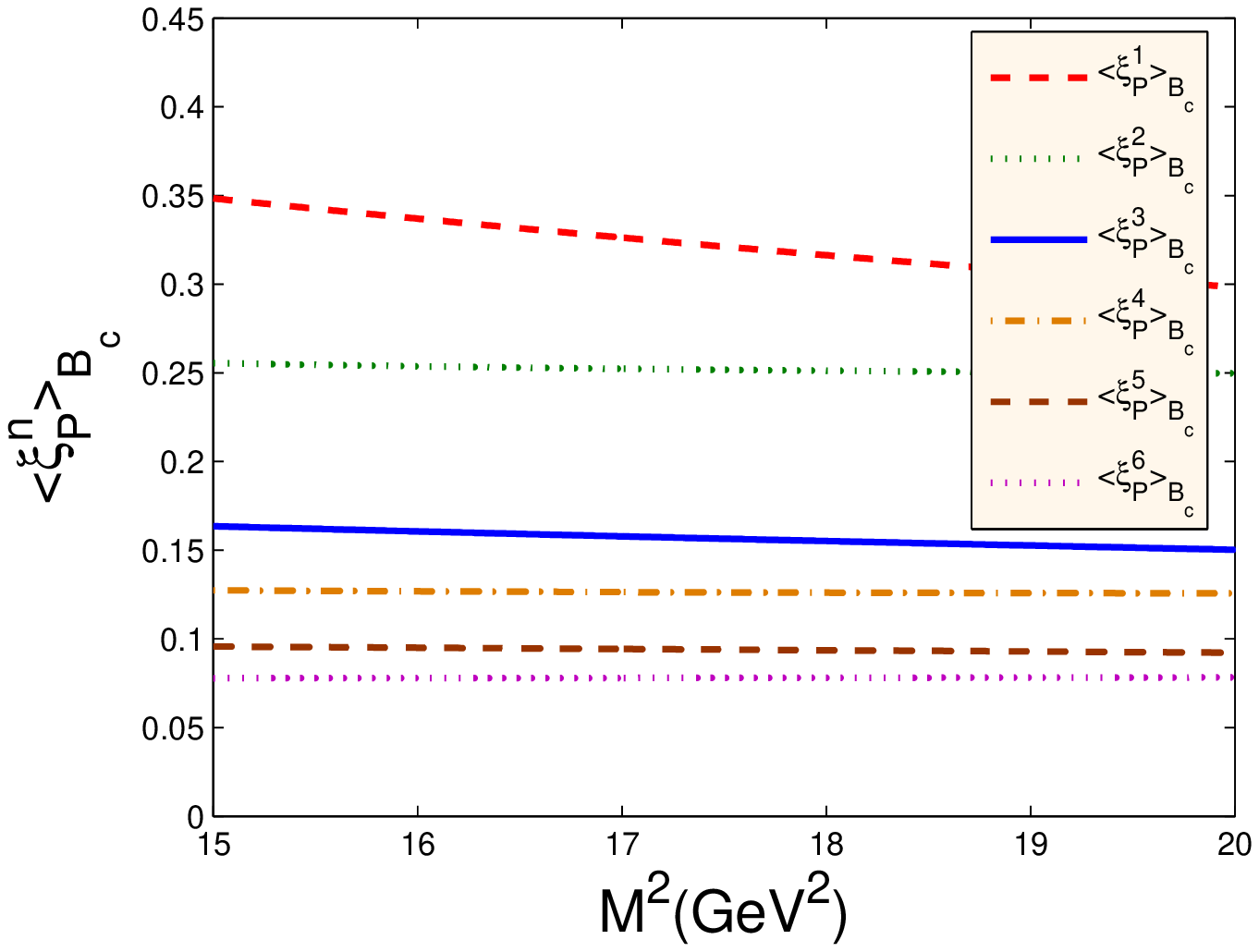}
\includegraphics[width=0.32\textwidth]{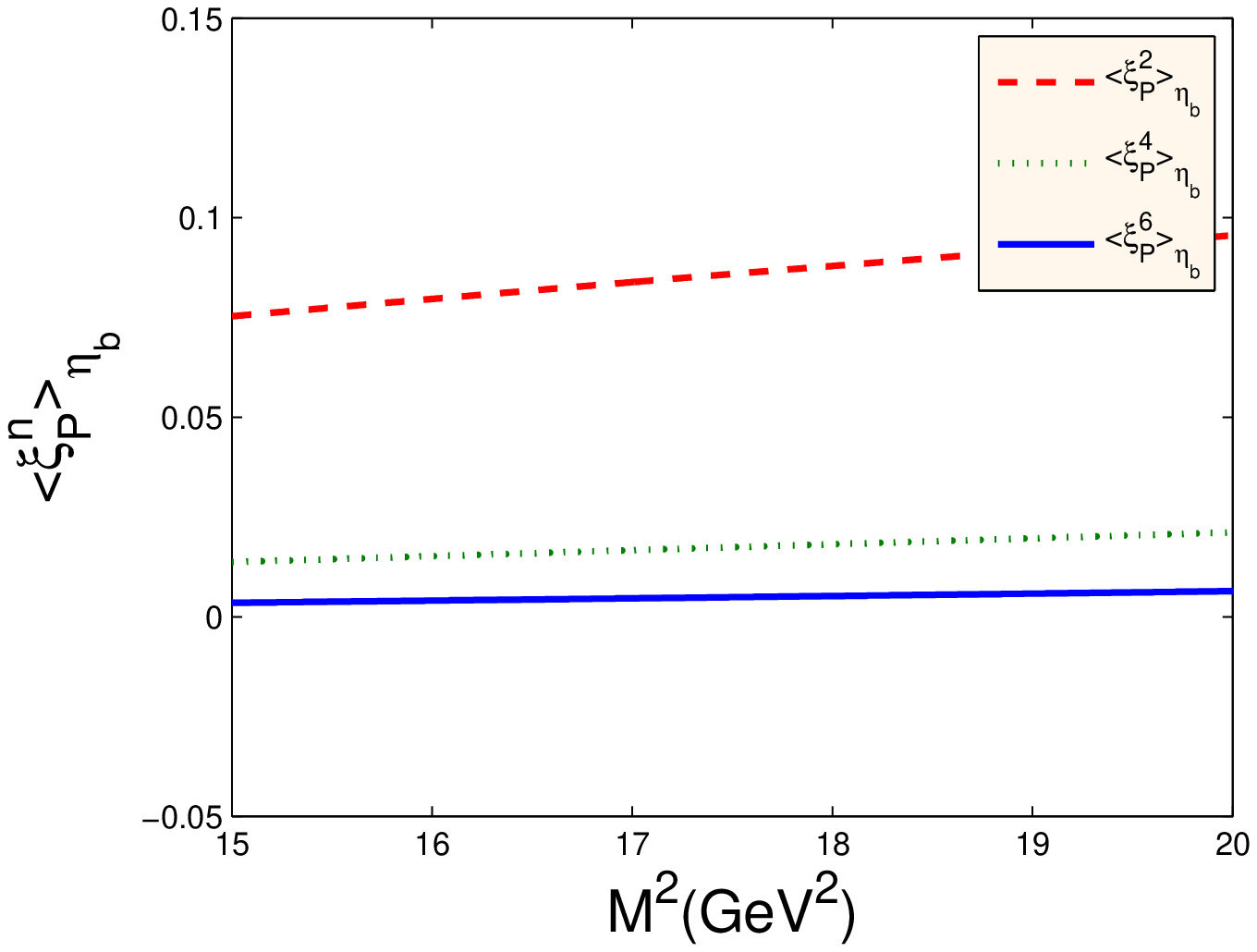}
\caption{The moments $\left<\xi^n_P\right>_{\rm HP}$ $(n\leq6)$ of the HP twist-3 DA $\phi^P_{3;\rm HP}$ versus the Borel parameter $M^2$, where all the input parameters are set to be their central values.}
\label{fxinps}
\end{figure*}

\begin{figure*}[tb]
\centering
\includegraphics[width=0.32\textwidth]{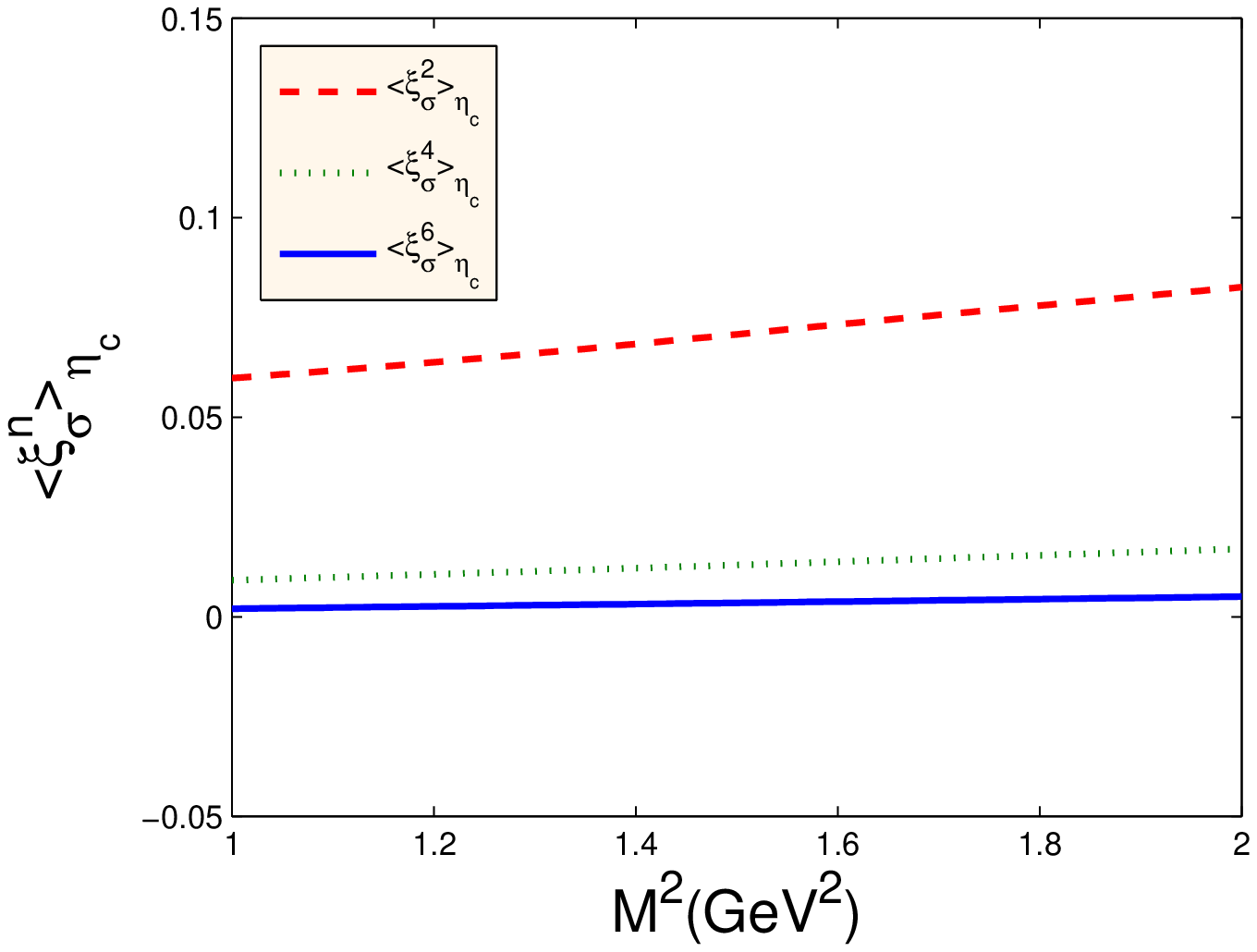}
\includegraphics[width=0.32\textwidth]{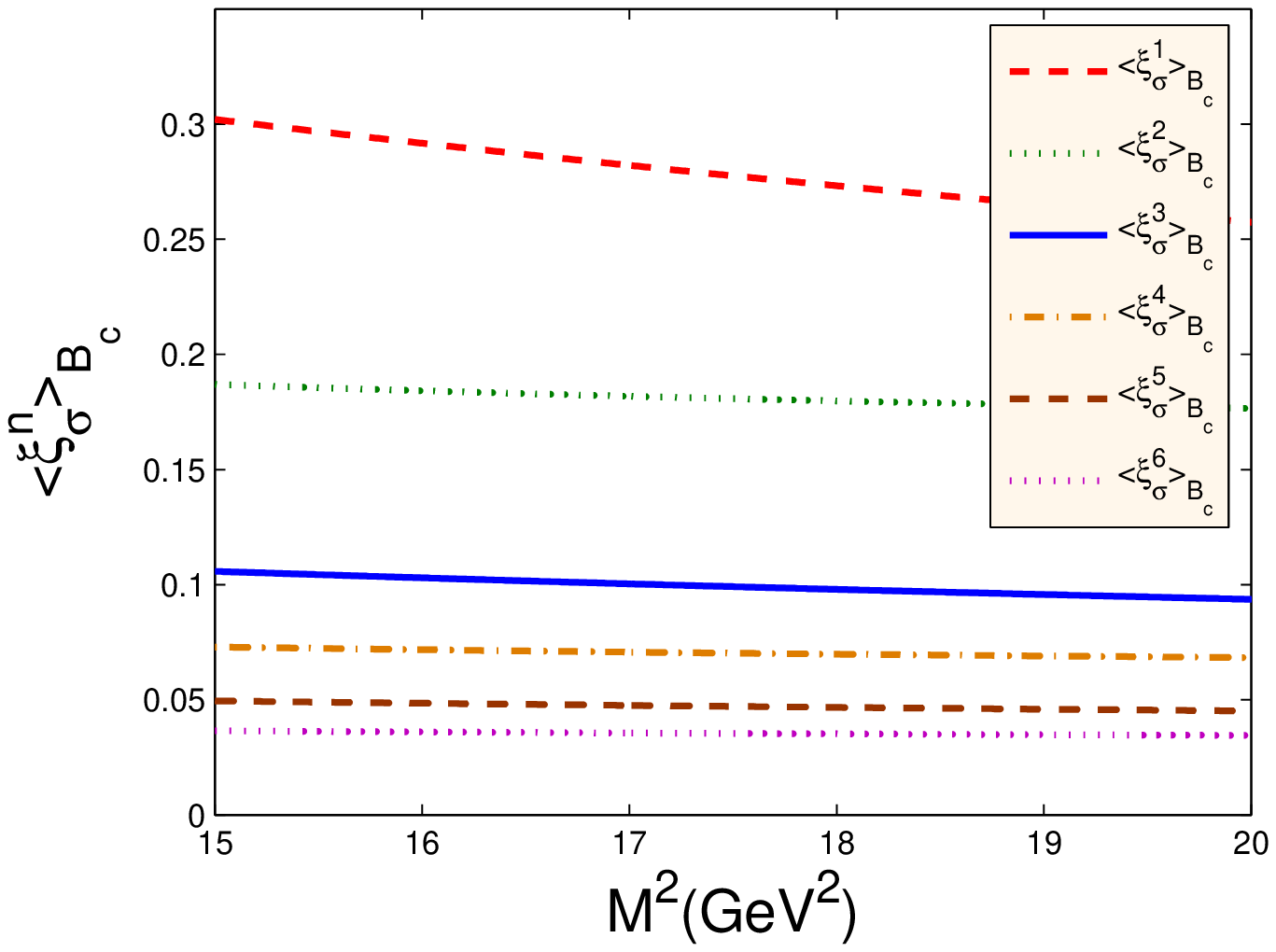}
\includegraphics[width=0.32\textwidth]{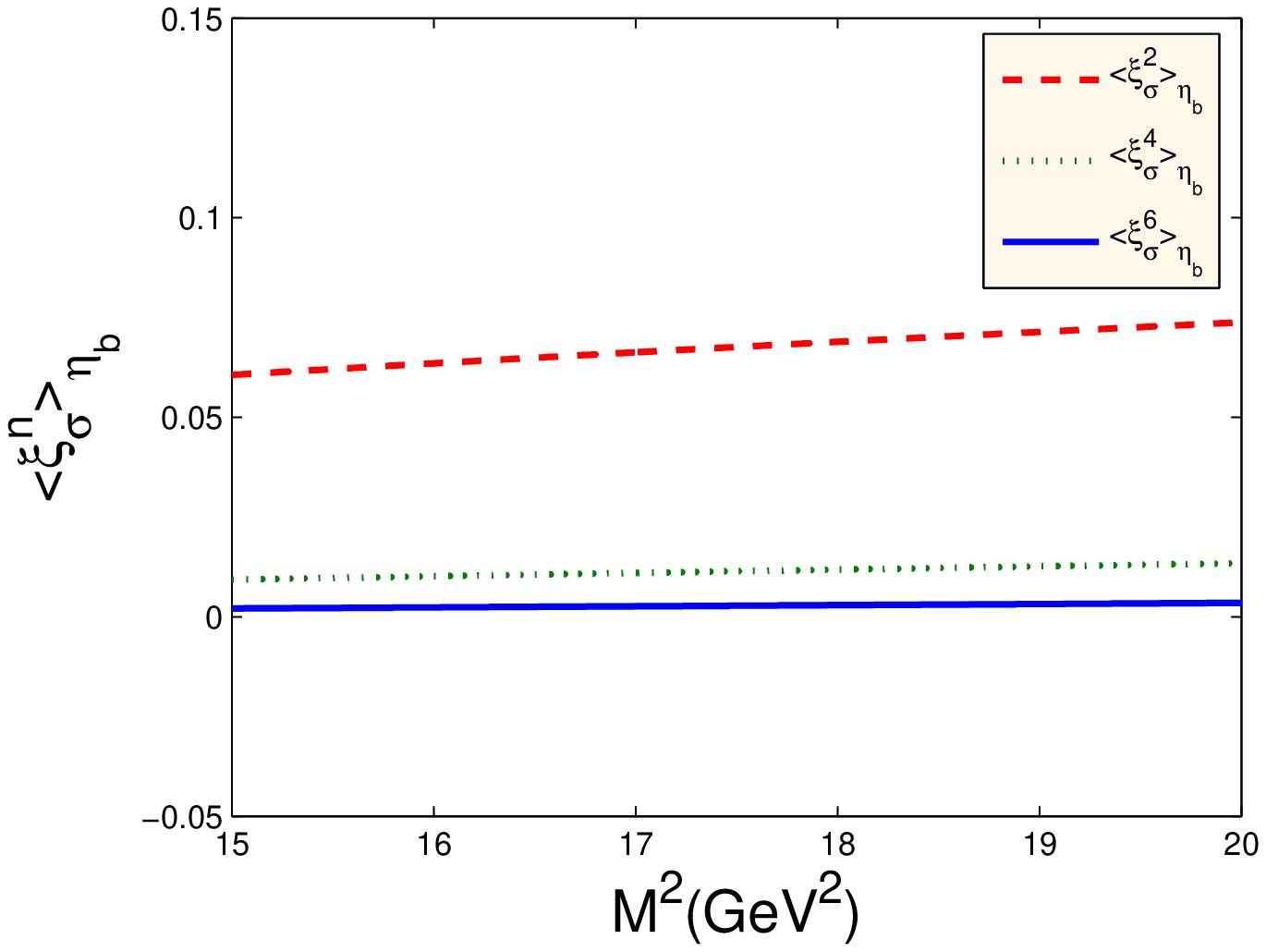}
\caption{The moments $\left<\xi^n_\sigma\right>_{\rm HP}$ $(n\leq6)$ of the HP twist-3 DA $\phi^\sigma_{3;\rm HP}$ versus the Borel parameter $M^2$, where all the input parameters are set to be their central values.}
\label{fxinpt}
\end{figure*}

\begin{table}[htb]
\caption{The HP twist-3 DA moments $\left<\xi^n_P\right>_{\rm HP}$ up to $6_{\rm th}$-order. The errors are squared average of those from all input parameters, such as the Borel parameter, the condensates and the bound state parameters. The scale $\mu$ is set to be $\bar{m}_c(\bar{m}_c)$ for $\eta_c$ and $\bar{m}_b(\bar{m}_b)$ for $B_c$ and $\eta_b$. }
\begin{tabular}{ c | c c c c }
\hline
~~ & ~$\eta_c(\bar{m}_c(\bar{m}_c))$~ & ~$B_c(\bar{m}_b(\bar{m}_b))$~ & ~$\eta_b(\bar{m}_b(\bar{m}_b))$~ \\
\hline
~$\left<\xi^1_P\right>_{\rm HP}$~& ~$       0       $~ & ~$0.323 \pm 0.025$~ & ~0~ \\
~$\left<\xi^2_P\right>_{\rm HP}$~& ~$0.096 \pm 0.015$~ & ~$0.253 \pm 0.004$~ & ~$0.086 \pm 0.010$~ \\
~$\left<\xi^3_P\right>_{\rm HP}$~& ~$       0       $~ & ~$0.157 \pm 0.007$~ & ~0~ \\
~$\left<\xi^4_P\right>_{\rm HP}$~& ~$0.023 \pm 0.007$~ & ~$0.127 \pm 0.002$~ & ~$0.017 \pm 0.004$~ \\
~$\left<\xi^5_P\right>_{\rm HP}$~& ~$       0       $~ & ~$0.094 \pm 0.003$~ & ~0~ \\
~$\left<\xi^6_P\right>_{\rm HP}$~& ~$0.008 \pm 0.003$~ & ~$0.078 \pm 0.002$~ & ~$0.005 \pm 0.001$~ \\
\hline
\end{tabular}
\label{txinps}
\end{table}

\begin{table}[htb]
\caption{The HP twist-3 DA moments $\left<\xi^n_\sigma\right>_{\rm HP}$ up to $6_{\rm th}$-order. The errors are squared average of those from all input parameters, such as the Borel parameter, the condensates and the bound state parameters. The scale $\mu$ is set to be $\bar{m}_c(\bar{m}_c)$ for $\eta_c$ and $\bar{m}_b(\bar{m}_b)$ for $B_c$ and $\eta_b$.
}
\begin{tabular}{ c | c c c c }
\hline
~~ & ~$\eta_c(\bar{m}_c(\bar{m}_c))$~ & ~$B_c(\bar{m}_b(\bar{m}_b))$~ & ~$\eta_b(\bar{m}_b(\bar{m}_b))$~ \\
\hline
~$\left<\xi^1_\sigma\right>_{\rm HP}$~& ~$       0       $~ & ~$0.279 \pm 0.023$~ & ~0~ \\
~$\left<\xi^2_\sigma\right>_{\rm HP}$~& ~$0.074 \pm 0.012$~ & ~$0.182 \pm 0.005$~ & ~$0.067 \pm 0.007$~ \\
~$\left<\xi^3_\sigma\right>_{\rm HP}$~& ~$       0       $~ & ~$0.100 \pm 0.006$~ & ~0~ \\
~$\left<\xi^4_\sigma\right>_{\rm HP}$~& ~$0.015 \pm 0.004$~ & ~$0.071 \pm 0.003$~ & ~$0.011 \pm 0.002$~ \\
~$\left<\xi^5_\sigma\right>_{\rm HP}$~& ~$       0       $~ & ~$0.047 \pm 0.002$~ & ~0~ \\
~$\left<\xi^6_\sigma\right>_{\rm HP}$~& ~$0.005 \pm 0.002$~ & ~$0.036 \pm 0.001$~ & ~$0.003 \pm 0.001$~ \\
\hline
\end{tabular}
\label{txinpt}
\end{table}

We first discuss the properties of the first several moments of the HP twist-3 DA $\phi^P_{3;\rm HP}$ and $\phi^\sigma_{3;\rm HP}$. Figs.(\ref{fxinps}, \ref{fxinpt}) show the stability of those moments versus the Borel parameter $M^2$, where all input parameters are set to be their central values. Tables \ref{txinps} and \ref{txinpt} display the moments $\left<\xi^n_P\right>_{\rm HP}$ and $\left<\xi^n_\sigma\right>_{\rm HP}$ up to $6_{\rm th}$-order, in which the scale $\mu$ is set to be $\bar{m}_c(\bar{m}_c)$ for $\eta_c$ and $\bar{m}_b(\bar{m}_b)$ for $B_c$ and $\eta_b$, respectively. In Tables \ref{txinps} and \ref{txinpt}, the errors are squared average of those from all input parameters, such as the Borel parameter, the vacuum condensates and the heavy quark masses. All twist-3 DA moments follow the same trend that a smaller moment is achieved when its order is larger, which explains why people usually only takes into account the first several DA moments to do the discussion. This trend is much more obvious for the cases of $\eta_c$ and $\eta_b$, i.e.
\begin{displaymath}
\left<\xi^2_P\right>_{\eta_c (\eta_b)} : \left<\xi^4_P\right>_{\eta_c (\eta_b)}: \left<\xi^6_P\right>_{\eta_c (\eta_b)} \simeq 1:0.2:0.08(0.06)
\end{displaymath}
and
\begin{displaymath}
\left<\xi^2_\sigma\right>_{\eta_c (\eta_b)} : \left<\xi^4_\sigma\right>_{\eta_c (\eta_b)}: \left<\xi^6_\sigma\right>_{\eta_c (\eta_b)} \simeq 1:0.2:0.07(0.04).
\end{displaymath}

\subsection{Properties of the HP twist-3 DA}

\begin{table}[htb]
\caption{The determined model parameters of the HP twist-3 DA $\phi^P_{3;\rm HP}$ at the scale $\mu = \bar{m}_b(\bar{m}_b)$.}
\begin{tabular}{ c | c c c c }
\hline
~{\rm DA}~ & ~$\phi^P_{3;\eta_c}$~ & ~$\phi^P_{3;B_c}$~ & ~$\phi^P_{3;\eta_b}$~ \\
\hline
~$A^P_{\rm HP}({\rm GeV}^{-1})$ ~& ~$2.616$~ & ~$3.101$~ & ~$9.987$~ \\
~$\beta_{\rm HP}^P({\rm GeV})$  ~& ~$3.106$~ & ~$3.626$~ & ~$2.962$~ \\
~$B^{{\rm HP},P}_1$             ~& ~$0$~      & ~$1.378$~ & ~$0$~ \\
~$B^{{\rm HP},P}_2$             ~& ~$0.780$~ & ~$0.744$~ & ~$-1.858$~ \\
~$B^{{\rm HP},P}_3$             ~& ~$0$~      & ~$0.757$~ & ~$0$~ \\
~$B^{{\rm HP},P}_4$             ~& ~$1.920$~ & ~$0.832$~ & ~$0.352$~ \\
~$B^{{\rm HP},P}_5$             ~& ~$0$~      & ~$0.378$~ & ~$0$~ \\
~$B^{{\rm HP},P}_6$             ~& ~$1.021$~ & ~$0.266$~ & ~$0.081$~ \\
\hline
\end{tabular}
\label{tmpps}
\end{table}

\begin{table}[htb]
\caption{The determined model parameters of the HP twist-3 DA $\phi^\sigma_{3;\rm HP}$ at the scale $\mu = \bar{m}_b(\bar{m}_b)$.}
\begin{tabular}{ c | c c c c }
\hline
~{\rm DA}~ & ~$\phi^\sigma_{3;\eta_c}$~ & ~$\phi^\sigma_{3;B_c}$~ & ~$\phi^\sigma_{3;\eta_b}$~ \\
\hline
~$A^\sigma_{\rm HP}({\rm GeV}^{-1})$~& ~$2.508$~  & ~$3.276$~ & ~$17.235$~ \\
~$\beta_{\rm HP}^\sigma({\rm GeV})$ ~& ~$2.799$~  & ~$3.243$~ & ~$2.773$~ \\
~$B^{{\rm HP},\sigma}_1$            ~& ~$0$~       & ~$0.372$~ & ~$0$~ \\
~$B^{{\rm HP},\sigma}_2$            ~& ~$-0.257$~ & ~$-0.115$~ & ~$-0.359$~ \\
~$B^{{\rm HP},\sigma}_3$            ~& ~$0$~       & ~$-0.085$~ & ~$0$~ \\
~$B^{{\rm HP},\sigma}_4$            ~& ~$0.095$~  & ~$0.009$~ & ~$0.092$~ \\
~$B^{{\rm HP},\sigma}_5$            ~& ~$0$~       & ~$0.002$~ & ~$0$~ \\
~$B^{{\rm HP},\sigma}_6$            ~& ~$-0.010$~ & ~$-0.003$~ & ~$-0.012$~ \\
\hline
\end{tabular}
\label{tmppt}
\end{table}

By using the moments $\left<\xi^n_P\right>_{\rm HP}$ and $\left<\xi^n_\sigma\right>_{\rm HP}$ presented in Tables \ref{txinps} and \ref{txinpt}, we are ready to fix the input parameters for the HP twist-3 DA $\phi^P_{3;\rm HP}$ and $\phi^\sigma_{3;\rm HP}$. The results of those parameters at the scale $\mu = \bar{m}_b(\bar{m}_b)$ are presented in Tables \ref{tmpps} and \ref{tmppt}, where all input parameters including the moments $\left<\xi^n_P\right>_{\rm HP}$ and $\left<\xi^n_\sigma\right>_{\rm HP}$ are set to be their central values. One can get the DA model parameters at any other scales via the evolution equation of the HP DA~\cite{HP_T2DA_zhong, EE}~\footnote{Another equivalent approach is to evolute the moments $\left<\xi^n_P\right>_{\rm HP}$ and $\left<\xi^n_\sigma\right>_{\rm HP}$ listed in Tables \ref{txinps} and \ref{txinpt} to the required scale by using the renormalization group equation for the DA moments~\cite{RGE}, and then solve the constraints (\ref{norma_condi}, \ref{avera_value_k}, \ref{mom_t3ps}).}. For example, at the scale $\mu=\bar{m}_c(\bar{m}_c)$, we have $A^P_{\eta_c} = 542.074 {\rm GeV}^{-1},\ B^{{\eta_c},P}_2 = 1.329,\ B^{{\eta_c},P}_4 = 1.219,\ B^{{\eta_c},P}_6 = 0.382,\ \beta^P_{\eta_c} = 0.834 {\rm GeV}$ for $\phi^P_{3;\eta_c}$; and $A^\sigma_{\eta_c} = 736.146 {\rm GeV}^{-1},\ B^{{\eta_c},\sigma}_2 = 0.327,\ B^{{\eta_c},\sigma}_4 = 0.352,\ B^{{\eta_c},\sigma}_6 = 0.108,\ \beta^\sigma_{\eta_c} = 0.770 {\rm GeV}$ for $\phi^\sigma_{3;\eta_c}$.

\begin{figure}[tb]
\centering
\includegraphics[width=0.45\textwidth]{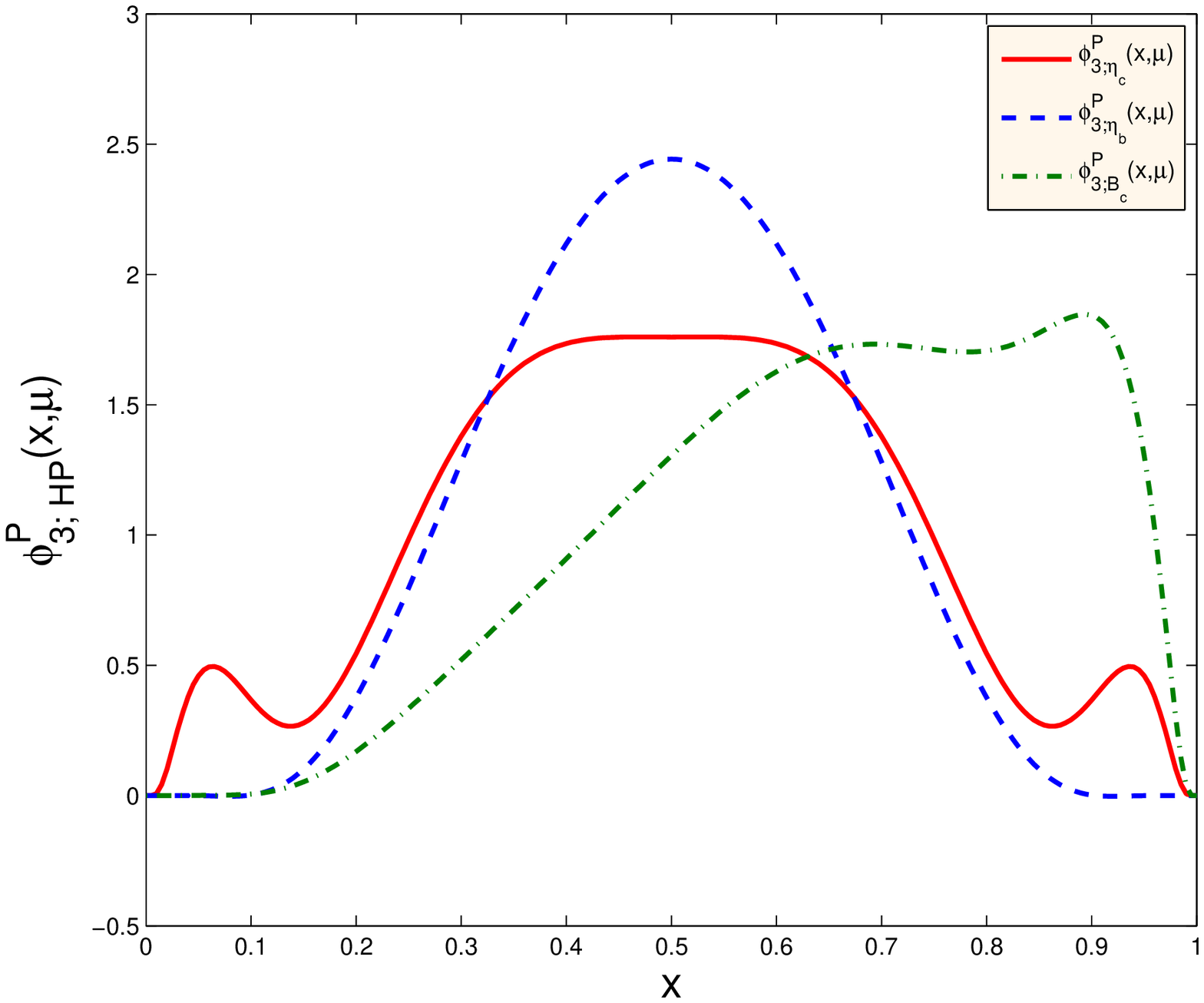}
\caption{The curves of the HP twist-3 DA $\phi^P_{3;\rm HP}(x,\mu)$ at the scale $\mu = \bar{m}_b(\bar{m}_b)$.}
\label{fphips}
\end{figure}

\begin{figure}[tb]
\centering
\includegraphics[width=0.45\textwidth]{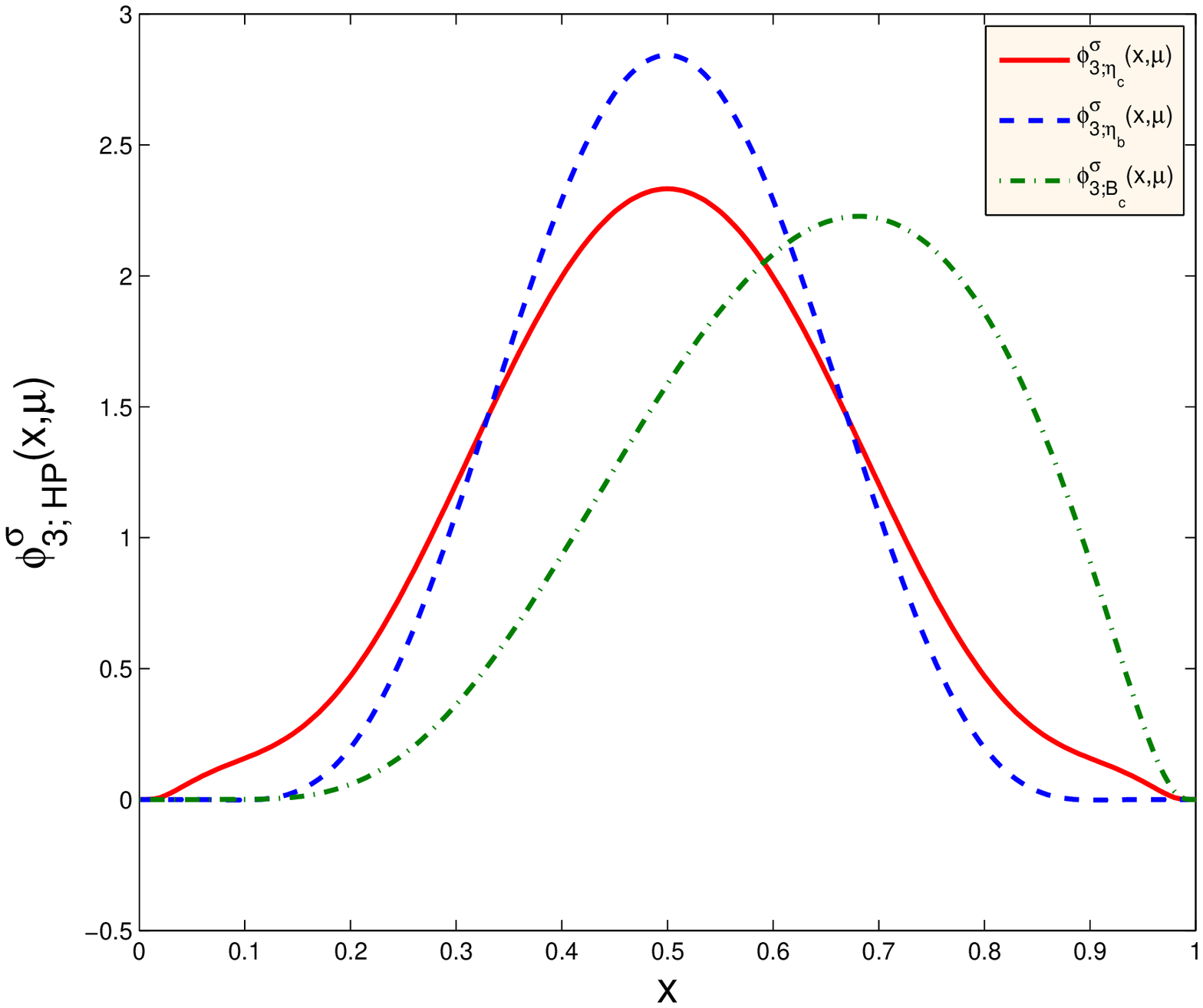}
\caption{The curves of the HP twist-3 DA $\phi^\sigma_{3;\rm HP}(x,\mu)$ at the scale $\mu = \bar{m}_b(\bar{m}_b)$.}
\label{fphipt}
\end{figure}

\begin{figure*}[tb]
\centering
\includegraphics[width=0.32\textwidth]{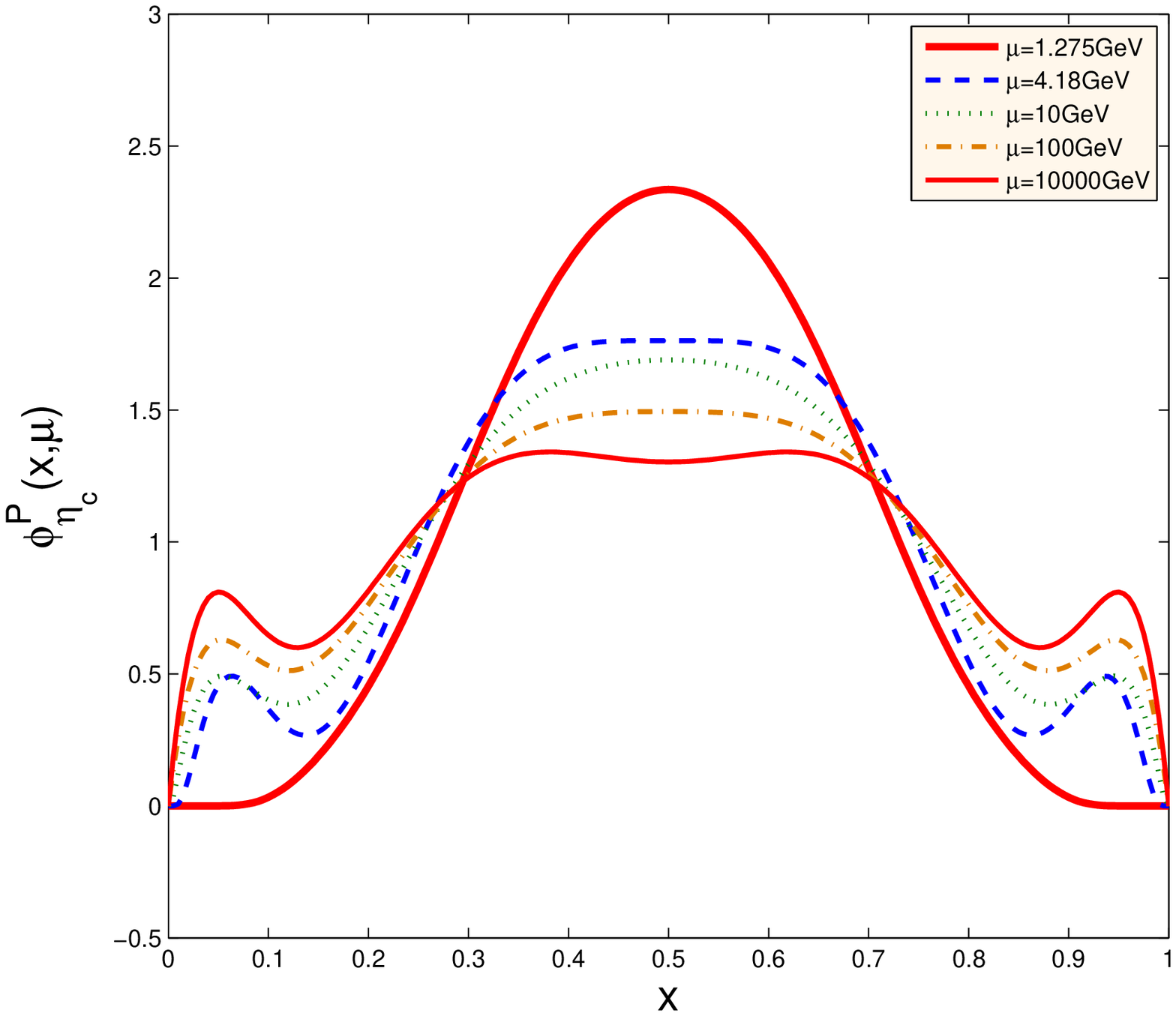}
\includegraphics[width=0.32\textwidth]{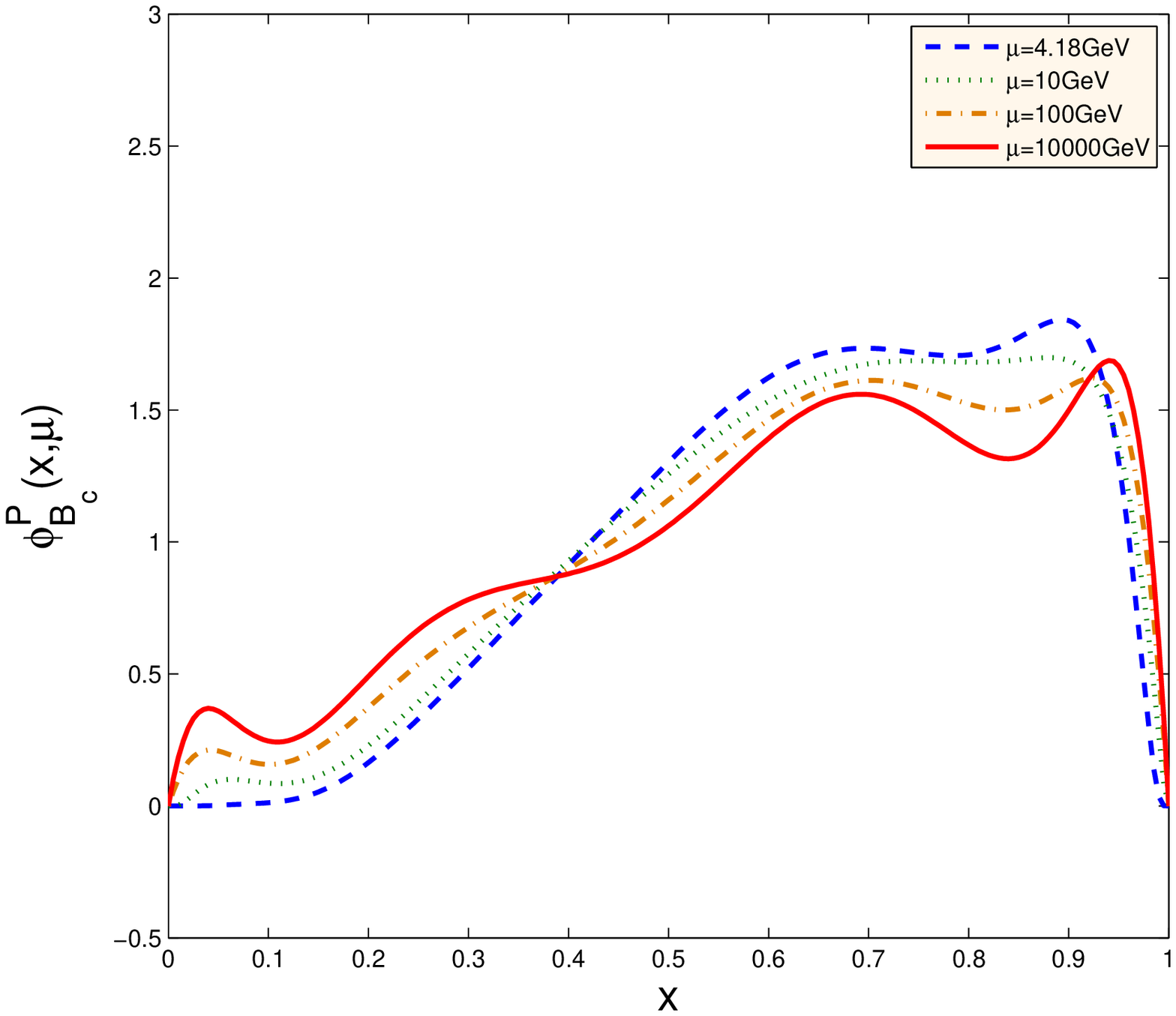}
\includegraphics[width=0.32\textwidth]{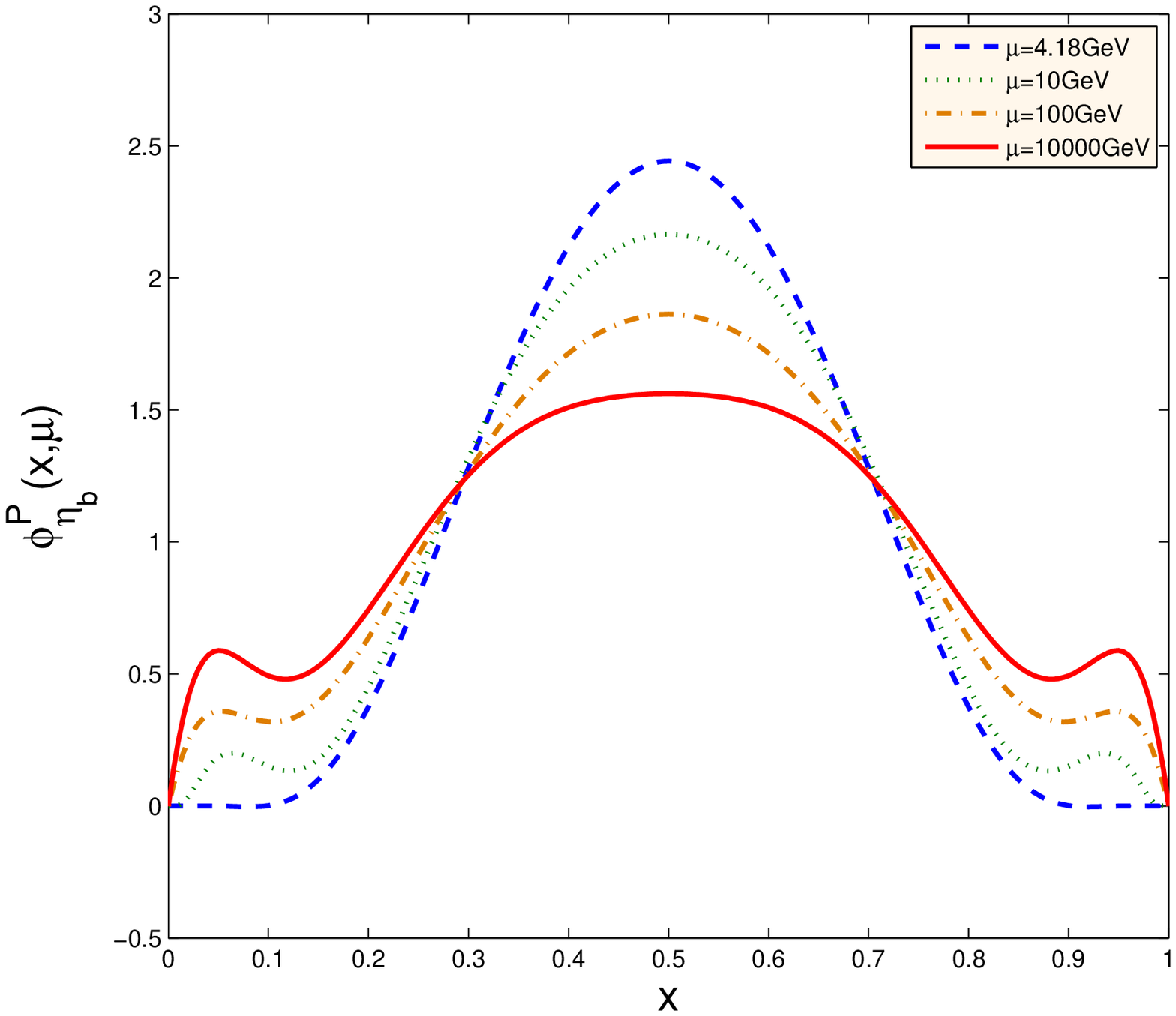}
\caption{The running of the HPs' twist-3 DA $\phi^P_{3;\rm HP}(x,\mu)$. The dashed, the dotted, the dash-dot and the solid lines are for $\mu = 4.18{\rm GeV}$, $10{\rm GeV}$, $100{\rm GeV}$ and $10000 \rm GeV$, respectively. Moreover, in the first figure the thick solid line is for $\mu = 1.275 \rm GeV$.}
\label{fDAPS_evolution}
\end{figure*}

\begin{figure*}[tb]
\centering
\includegraphics[width=0.32\textwidth]{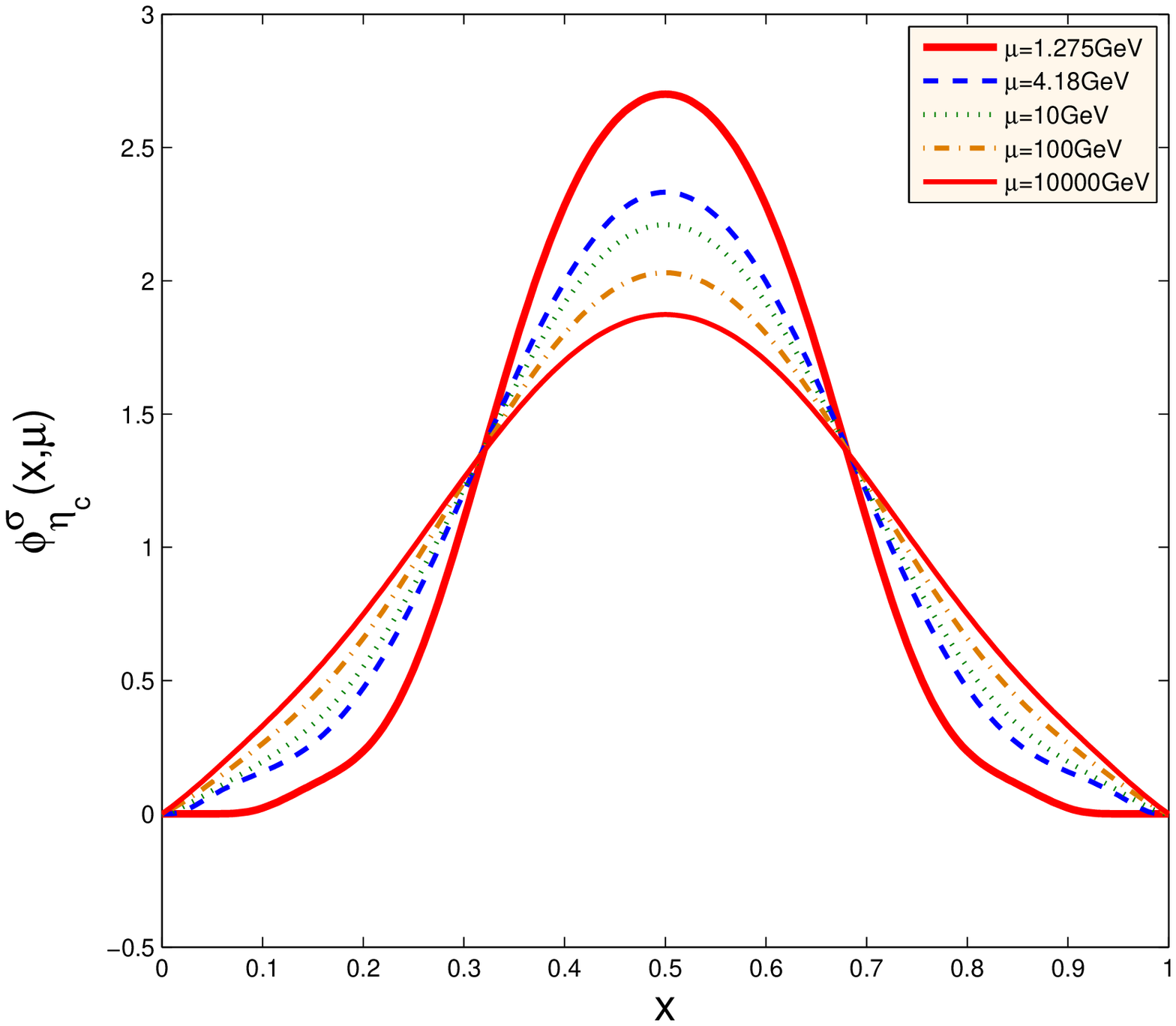}
\includegraphics[width=0.32\textwidth]{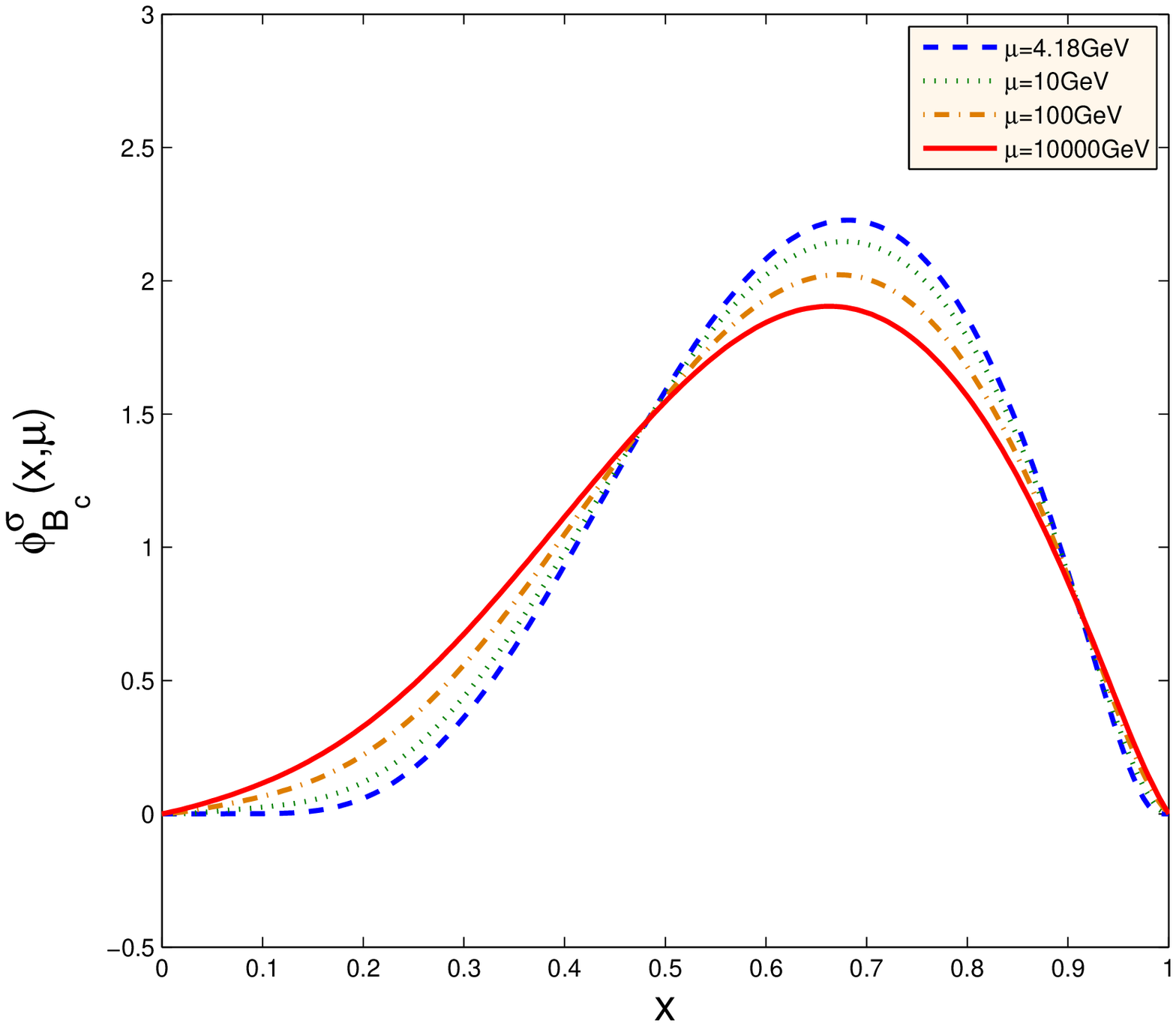}
\includegraphics[width=0.32\textwidth]{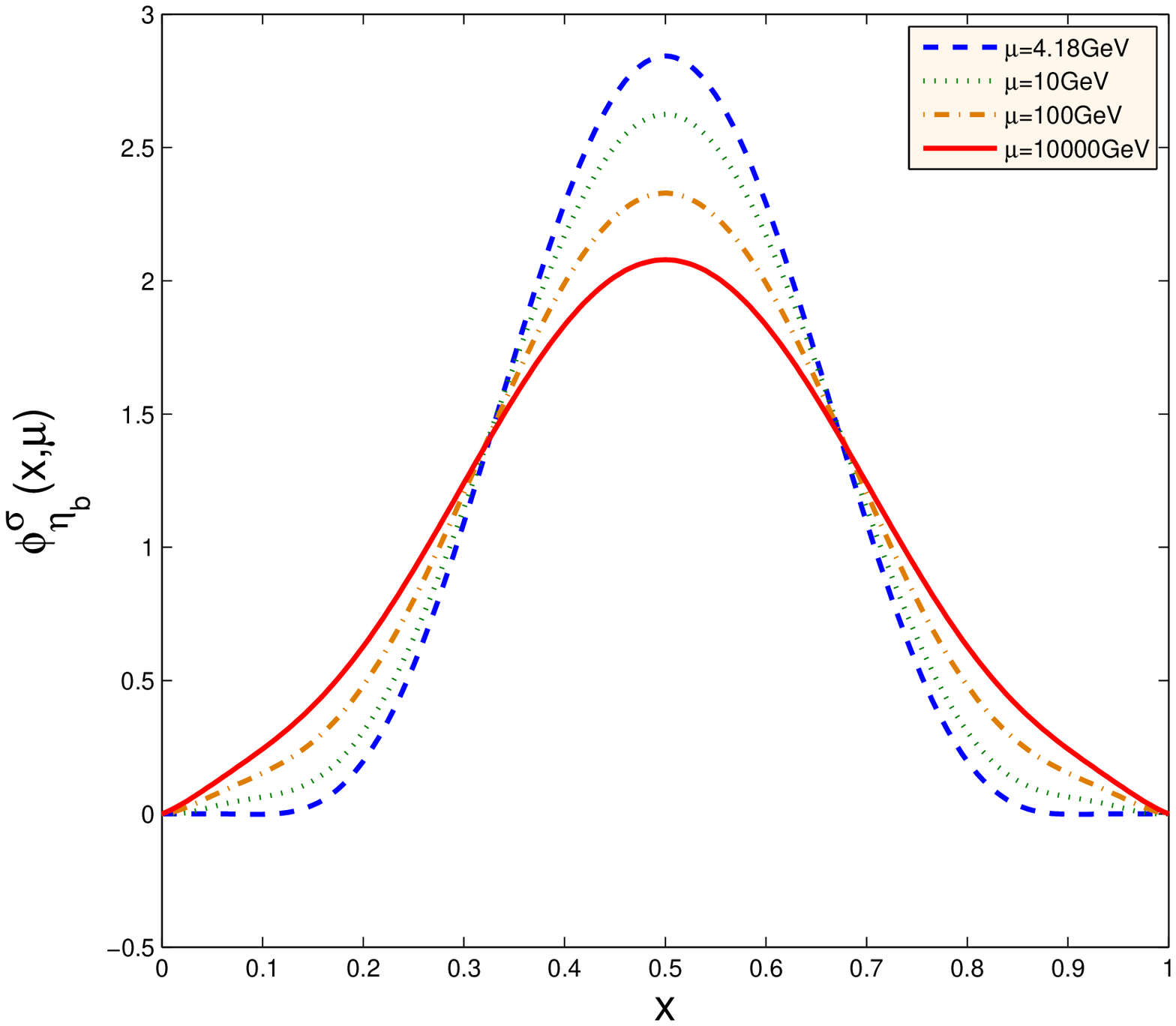}
\caption{The running of the HPs' twist-3 DA $\phi^\sigma_{3;\rm HP}(x,\mu)$. The dashed, the dotted, the dash-dot and the solid lines are for $\mu = 4.18{\rm GeV}$, $10{\rm GeV}$, $100{\rm GeV}$ and $10000 \rm GeV$, respectively. Moreover, in the first figure the thick solid line is for $\mu = 1.275 \rm GeV$.}
\label{fDAPT_evolution}
\end{figure*}

We present the twist-3 DAs $\phi^P_{3;\rm HP}$ and $\phi^\sigma_{3;\rm HP}$ at the scale $\mu = \bar{m}_b(\bar{m}_b)$ in Figs.(\ref{fphips}, \ref{fphipt}). As required, the $\eta_c$ and $\eta_b$ twist-3 DAs are symmetric and the $B_c$ twist-3 DAs are asymmetric. To show more clearly how those twist-3 DAs change with the scale changes, we present the twist-3 DAs $\phi^P_{3;\rm HP}(x,\mu)$ and $\phi^\sigma_{3;\rm HP}(x,\mu)$ under several typical scales in Figs.(\ref{fDAPS_evolution}, \ref{fDAPT_evolution}), where the dashed, the dotted, the dash-dot and the solid lines are for $\mu = \bar{m}_b(\bar{m}_b)=4.18{\rm GeV}$, $10{\rm GeV}$, $100{\rm GeV}$ and $10000 \rm GeV$, respectively. In those figures, we also present the results for $\phi^P_{3;\eta_c}$ and $\phi^\sigma_{3;\eta_c}$ at the scale $\mu = \bar{m}_c(\bar{m}_c)=1.275 \rm GeV$. It is found that $\phi^\sigma_{3;\eta_c}$ and $\phi^\sigma_{3;\eta_b}$ are close in shape, both of which tend to the asymptotic form $6x(1-x)$ when $\mu \to \infty$~\cite{EE}. $\phi^P_{3;\eta_c}(x,\bar{m}_c(\bar{m}_c))$ and $\phi^P_{3;\eta_b}(x,\bar{m}_b(\bar{m}_b))$ are also close in shape.
When $\phi^P_{3;\eta_c}$ and $\phi^P_{3;\eta_b}$ run to high scales, one observes a humped behavior near the end-point region $x\to 0$ or $1$~\footnote{Because the $c$-quark is lighter than the $b$-quark, $\phi^P_{3;\eta_c}$ shows the humped behavior more quickly with the increment of scale than $\phi^P_{3;\eta_b}$ and is more transparent at the same scale.}. Such humped behavior can be explained as a combination effect of the asymptotic behavior $\phi^{P}_{3;\eta_{c(b)}}(x,\mu\to\infty)=1$ and the end-point suppression as indicated by Eq.(\ref{DA_model_P}).

\section{The application of the HP twist-3 DA to the TFF $f^{B_c\to\eta_c}_+(q^2)$}

In the literature, in order to suppress the contributions from the higher-twist DA, which are uncertain and less-known, people have suggested to use a chiral correlator other than the conventional correlator to do the light-cone sum rules (LCSR) calculation~\cite{ccc}. In addition to the $B\to\pi$ transition~\cite{B_PI1, B_PI2, B_PI3}, this method has also been applied to other transitions such as the $B(D)\to \pi$ transition~\cite{BD_PI}, the $B\to K$ transition~\cite{B_k, B_P}, the $B(B_s,B_c)\to P(V)$ transition~\cite{B_PV, B_Ks, B_rho}, the $B\to D$ transition~\cite{B_D}, the $B\to S$ transition~\cite{B_S} and etc. In Ref.\cite{HP_T2DA_zhong}, we have adopted the chiral correlator for the TFF $f^{B_c\to\eta_c}_+(q^2)$, which does only contain the $\eta_c$ leading-twist DA in the LCSR. In this section, as an application of the present suggested HP twist-3 DA model, we shall calculate the TFF $f^{B_c\to\eta_c}_+(q^2)$ by using the conventional correlator to do our LCSR discussion, in which both the twist-2 and twist-3 terms have been kept.

\subsection{Properties of the LCSR prediction of $f^{B_c\to\eta_c}_+(q^2)$ with the conventional correlator}

The $B_c\to\eta_c$ TFF $f^{B_c\to\eta_c}_+(q^2)$ is defined by the following matrix element
\begin{eqnarray}
\left<\eta_c(p)\left|\bar{c}\gamma_\mu b\right|B_c(p+q)\right> &=& 2f^{B_c\to\eta_c}_+(q^2) p_\mu + \left[ f^{B_c\to\eta_c}_+(q^2) \right.\nonumber\\
&+& \left. f^{B_c\to\eta_c}_-(q^2) \right] q_\mu. \label{tff_defin}
\end{eqnarray}
By using the conventional correlator
\begin{eqnarray} &&
\Pi_\mu(p,q) \nonumber\\&&
= i \int d^4x e^{iq\cdot x} \left<\eta_c(p)\left| \bar{c}(x) \gamma_\mu b(x), (m_b+m_c) \bar{b}(0) i\gamma_5 c(0) \right|0\right> \nonumber\\&&
= F\left[ (p+q)^2 \right] p_\mu + q_\mu \textrm{terms}, \label{cor_tff}
\end{eqnarray}
and following the standard procedure of the LCSR approach~\cite{LCSR1, LCSR2, LCSR3}, one can obtain the LCSR of $f^{B_c\to\eta_c}_+(q^2)$, which can be formulated as a function of $\eta_c$ twist-2 and twist-3 DAs, i.e.
\begin{eqnarray}
f^{B_c\to\eta_c}_+(q^2) &=& \frac{m_b (m_b+m_c) f_{\eta_c}}{2m_{B_c}^2 f_{B_c}} e^{m_{B_c}^2/M^2} \nonumber\\
&\times& \int^1_{\Delta} du e^{- \frac{m_b^2 - \bar{u}q^2 + u\bar{u}m_{\eta_c}^2}{uM^2}} \left\{ \frac{\phi_{2;\eta_c}(u)}{u} + \frac{\mu_{\eta_c}}{m_b} \right. \nonumber\\
&\times& \left[ \phi_{3;\eta_c}^P(u) + \frac{1}{6} \left( 1 - \frac{4m_c^2}{m_{\eta_c}^2} \right) \right. \nonumber\\
&\times& \left( \frac{2\phi_{3;\eta_c}^\sigma(u)}{u} + \frac{1}{m_b^2 - q^2 + u^2 m_{\eta_c}^2} \right. \nonumber\\
&\times& \left[ \frac{4u m_b^2 m_{\eta_c}^2}{m_b^2 - q^2 + u^2 m_{\eta_c}^2} \phi_{3;\eta_c}^\sigma(u) \right. \nonumber\\
&-& \left.\left.\left.\left. (m_b^2 + q^2 - u^2 m_{\eta_c}^2) \frac{d\phi_{3;\eta_c}^\sigma(u)}{du} \right] \right) \right] \right\}, \label{sr_TFF}
\end{eqnarray}
where $\phi_{2;\eta_c}(u)$ is the leading-twist DA of $\eta_c$, $\bar{u} = 1-u$, and
\begin{eqnarray}
\Delta &=& \left[\sqrt{(s_0 - q^2 - m_{\eta_c}^2)^2 + 4m_{\eta_c}^2 (m_b^2 - q^2)} \right. \nonumber\\
&-& \left. (s_0 - q^2 - m_{\eta_c}^2)\right] / \left(2m_{\eta_c}^2\right). \label{del}
\end{eqnarray}

\begin{table}[htb]
\caption{A comparison of the $B_c\to\eta_c$ TFF $f^{B_c\to\eta_c}_+(0)$ under various approaches, i.e. the LCSR, the QCD sum rules (SR), the quark model (QM), the Bauer-Stech-Wirbel (BSW) framework, the perturbative QCD (pQCD) approach and the non-relativistic QCD (NRQCD) appraoch.}
\begin{tabular}{ c | c c }
\hline
~Approach~ & ~$f^{B_c\to\eta_c}_+(0)$~ & ~Ref.~ \\
\hline
~LCSR~& ~$0.674 \pm 0.066$~  & ~This work~ \\
~~& ~$0.612^{+0.053}_{-0.052}$~ & ~\cite{HP_T2DA_zhong}~ \\
\hline
~QCD SR~ & ~$0.66$~ & ~\cite{kiselev}~ \\
\hline
~QM~& ~$0.61$~  & ~\cite{BC_QM3}~ \\
~~& ~$0.49^{+0.01}$~  & ~\cite{BC_QM4}~ \\
~~& ~$0.61^{+0.03+0.01}_{-0.04-0.01}$~  & ~\cite{BRQM12}~ \\
~~& ~$0.47$~  & ~\cite{BRQM21}~ \\
\hline
~BSW~& ~$0.58^{+0.02}_{-0.01}$~  & ~\cite{BC_BSW}~ \\
\hline
~LO pQCD~& ~$0.48 \pm 0.06 \pm 0.01$~  & ~\cite{BRPQCD}~ \\
\hline
~NLO NRQCD~& ~$1.65$~  & ~\cite{TFF_PMC}~ \\
           & ~$1.28$~  & ~\cite{TFF_Qiao} \footnote{This value is obtained by using the formulas of Ref.\cite{TFF_Qiao} but with the same parameter values of Ref.\cite{TFF_PMC}.}~ \\
\hline
\end{tabular}
\label{ttff}
\end{table}

To do the numerical calculation, the $\eta_c$ twist-2 DA constructed in Ref.\cite{HP_T2DA_zhong} and the $\eta_c$ twist-3 DA as shown by Eq.(\ref{DA_model_P}) are adopted. As for its continuum threshold parameter and the Borel window, we take them to be $s_0 = 42 \textrm{GeV}^2$ and $M^2 = (20 - 35) \textrm{GeV}^2$. At the maximum recoil region, we obtain
\begin{eqnarray}
f_+^{B_c \to \eta_c}(0) = 0.674 \pm 0.066,   \label{tff0}
\end{eqnarray}
where all the theoretical uncertainties such as those from the Borel parameter and the bound state parameters have been added up in quadrature. A comparison of $f^{B_c\to\eta_c}_+(0)$ under various approaches~\cite{HP_T2DA_zhong, kiselev, BC_QM3, BC_QM4, BRQM12, BRQM21, BC_BSW, BRPQCD, TFF_PMC, TFF_Qiao} is put in Table \ref{ttff}, where `LO' is the leading-order prediction and `NLO' is the next-to-leading order prediction. The predictions from the LCSR, the quark model and the Bauer-Stech-Wirbel framework are at the LO level, most of them are consistent with each other within reasonable theoretical errors. The NLO NRQCD predictions are much bigger~\cite{TFF_PMC, TFF_Qiao}, indicating the importance of the NLO-terms. Moreover, Ref.\cite{TFF_PMC} shows the necessity of a proper renormalization scale-setting, i.e. it is important to achieve a reliable lower-order prediction.

By using the branching ratio formulas listed in Ref.\cite{HP_T2DA_zhong} but with the present LCSR for $f^{B_c\to\eta_c}_+(q^2)$, we finally predict the branching ratio of $B_c \to \eta_c l \nu$ as
\begin{equation}
Br(B_c \to \eta_c l \nu) = (9.31^{+2.27}_{-2.01})  \times 10^{-3},
\end{equation}
where the error of the branching ratio originates mainly from the TFF $f^{B_c\to\eta_c}_+(q^2)$, the CKM matrix element $|V_{cb}|$ and the $B_c$ meson lifetime $\tau_{B_c}$. The advantage of LCSR prediction in comparison to the pQCD ones lies in that the LCSRs is valid in a broader region $0< q^2 <m_b^2-2m_b\Lambda_{\rm QCD}\simeq 15 {\rm GeV}^2$, which is close to the allowable phase-space of $B_c \to \eta_cl\nu$, thus one may directly apply the LCSR prediction of $f^{B_c\to\eta_c}_+(q^2)$ to calculate the branching ratio $Br(B_c \to \eta_c l \nu)$; while, the pQCD prediction is only reliable in low $q^2$-region around the maximum recoil point, thus certain extrapolation must be applied, different choice of which shall introduce large extra error into the prediction.

\subsection{A comparison of the LCSRs under various correlators and the HP twist-3 DA}

\begin{figure}[htb]
\centering
\includegraphics[width=0.45\textwidth]{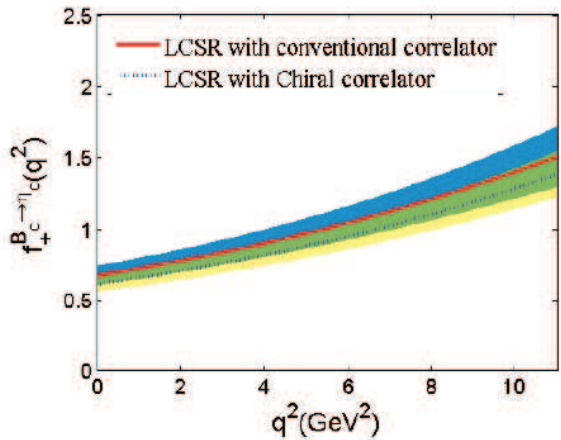}
\caption{A comparison of the LCSR predictions of $f^{B_c\to\eta_c}_+(q^2)$ under two different correlators, where the shaded hands are their uncertainties. The thicker shaded band stands for the LCSR with the chiral correlator and the lighter one stands for the LCSR with the conventional correlator. The dashed and solid lines are for their central values, respectively. }
\label{comLCSR}
\end{figure}

Table \ref{ttff} shows at the large recoil region, the two LCSR predictions and also the QCD sum rules prediction agree with each other within reasonable theoretical errors. A comparison of the LCSR predictions of $f^{B_c\to\eta_c}_+(q^2)$ under the conventional and the chiral correlators are shown in Fig.(\ref{comLCSR}), where the shaded hands are their uncertainties. Fig.(\ref{comLCSR}) shows the consistency of those two LCSR predictions is satisfied for all $q^2$-region. In Ref.\cite{HP_T2DA_zhong} a chiral correlator is adopted such that to suppress (or even eliminate) the unknown higher-twist contributions. In the present paper, we have shown that if one can construct a proper model for the twist-3 DA, one can also get a more accurate LCSR prediction.

\begin{figure}[htb]
\centering
\includegraphics[width=0.45\textwidth]{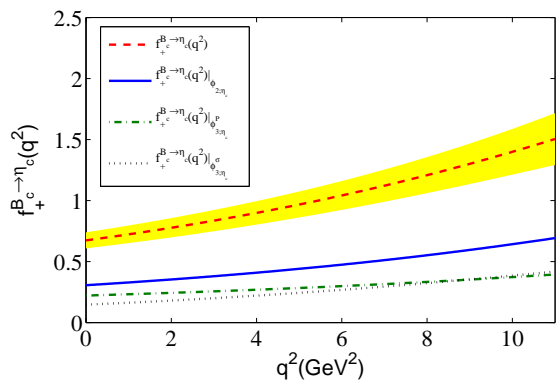}
\caption{The LCSR with the conventional correlator for the TFF $f^{B_c\to\eta_c}_+(q^2)$ versus $q^2$, where the shaded hand is its uncertainty. The solid, the dash-dot, the dotted and the dashed lines are for $\phi_{2;\eta_c}$, $\phi^{P}_{3;\eta_c}$, $\phi^{\sigma}_{3;\eta_c}$ and their sums, respectively.}
\label{ftffq}
\end{figure}

By using the LCSR with the conventional correlator, it is interesting to show what are the contributions for the TFF $f^{B_c\to\eta_c}_+(q^2)$ from different twist DA. We present the LCSR of $f_+^{B_c \to \eta_c}(q^2)$ versus $q^2$ in Fig.(\ref{ftffq}), where the contributions from $\phi_{2;\eta_c}$, $\phi^P_{3;\eta_c}$ and $\phi^\sigma_{3;\eta_c}$ are shown explicitly. The shaded hand shows its uncertainties, where all the errors have been added up in quadrature. At $q^2=0$, if setting all the parameters to be their central values, we have $f_+^{B_c \to \eta_c}(0)|_{\phi_{2;\eta_c}} = 0.306$, $f_+^{B_c \to \eta_c}(0)|_{\phi^P_{3;\eta_c}} = 0.221$ and $f_+^{B_c \to \eta_c}(0)|_{\phi^\sigma_{3;\eta_c}} = 0.147$, whose proportions in $f_+^{B_c \to \eta_c}(0)$ are $45\%$, $33\%$ and $22\%$, respectively. This shows the twist-3 contributions are large and important. For any other $q^2$-values, even though the twist-3 contribution become smaller, the conditions are similar. Moreover, when the value of $q^2$ increases, the twist-2 contribution increases faster, for a large enough $q^2$-value, e.g. $q^2>10\;{\rm GeV}^2$, we roughly have
\begin{eqnarray}
q^2 \frac{f_+^{B_c \to \eta_c}(q^2)|_{\phi^P_{3;\eta_c}}} {f_+^{B_c \to \eta_c}(q^2)|_{\phi_{2;\eta_c}}} & \rightsquigarrow & {\rm a\; flat\; line}, \\
q^2 \frac{f_+^{B_c \to \eta_c}(q^2)|_{\phi^P_{3;\eta_c}}} {f_+^{B_c \to \eta_c}(q^2)|_{\phi_{2;\eta_c}}} &\rightsquigarrow & {\rm a\; flat\; line},
\end{eqnarray}
indicating that the twist-3 contributions satisfy the usual $q^2$-suppression to the leading twist-2 contribution in large $q^2$-region. This in some sense shows the importance of the twist-3 DA with better end-point behavior. In fact, if taking $\phi^P_{3;\eta_c}$ to be its asymptotic form that has no end-point suppression, $\phi^P_{3;\eta_c}(x)\equiv1$, one cannot obtain such a $q^2$-suppression and the $\phi^P_{3;\eta_c}$ contribution shall be even dominant over the twist-2 one. This condition is similar to the case of $B\to\pi$ decays~\cite{ht1} and the pion form factor~\cite{pion1} , only if one has used a $\phi^P_{3;\pi}$ with better end-point behavior can one achieve the required power-suppressed twist-3 contribution.

\section{summary}

In this paper, based on the BHL-prescription, we have constructed a new model for the twist-3 DA $\phi^P_{3;\rm HP}$ and $\phi^\sigma_{3;\rm HP}$, which are for $\eta_c$, $B_c$ and $\eta_b$, respectively. To fit the parameters, we have calculated their moments $\left<\xi^n_P\right>_{\rm HP}$ and $\left<\xi^n_\sigma\right>_{\rm HP}$ by using the QCD SVZ sum rules under the framework of BFT. Tables \ref{txinps} and \ref{txinpt} display the values of those moments up to $6_{\rm th}$-order, which become smaller when they are at higher orders.

As an application of the constructed HP twist-3 DA, we have calculated the $B_c\to\eta_c$ TFF $f^{B_c\to\eta_c}_+$ by applying the LCSR with the conventional correlator. At the maximum recoil point, we obtain $f^{B_c\to\eta_c}_+(0) = 0.674 \pm 0.066$, which agrees with the predictions of other approaches also at the leading-order level. We have found that the net contributions from the $\eta_c$ twist-3 DA are large, which is up to $55\%$ for $f^{B_c\to\eta_c}_+(0)$. Such an importance of twist-3 DA can also be extended to other $q^2$-values as shown by Fig.(\ref{ftffq}). Furthermore, we have calculated the branching ratio of $B_c \to\eta_c l\nu$, $Br(B_c \to\eta_c l\nu) = \left( 9.31^{+2.27}_{-2.01} \right) \times 10^{-3}$, which is consistent with the previous LCSR prediction~\cite{HP_T2DA_zhong}. Being with a good end-point behavior due to the BHL-prescription, our present twist-3 contributions satisfy the usual $q^2$-suppression to the leading twist-2 contribution in large $q^2$-region.

The HP twist-3 DA are important inputs for analyzing the high-energy processes involving the HP, thus we think our present model shall have wide applications. \\

{\bf Acknowledgments}: This work was supported in part by the Natural Science Foundation of China under Grants No.11547015, No.11235005, No.11575110, No.11275280, and No.11547305, and by Fundamental Research Funds for the Central Universities under Grant No.CDJZR305513. \\

\appendix

\section*{Appendix: The expressions of ${\rm Im} I^{\rm PS,PT}_{\rm HP,pert}(s)$, $I^{\rm PS,PT}_{{\rm HP},\left<G^2\right>}(M^2)$ and $I^{\rm PS,PT}_{{\rm HP},\left<G^3\right>}(M^2)$}

In this appendix, we list the necessary expressions for deriving the sum rules of the moments $\left<\xi^n_P\right>_{\rm HP}$ and $\left<\xi^n_\sigma\right>_{\rm HP}$, which are
\begin{widetext}
\begin{eqnarray}
{\rm Im}I^{\rm PS}_{\rm HP,pert}(s) &=& \frac{3\bar{s}}{16\pi (n+1)} \left[ \left( \frac{-m_1^2 + m_2^2 + \bar{s}v}{s} \right)^{n+1} - \left( \frac{-m_1^2 + m_2^2 - \bar{s}v}{s} \right)^{n+1} \right], \\
I^{\rm PS}_{\rm HP,\left<G^2\right>}(M^2) &=& \left<\alpha_sG^2\right> \int^1_0 dx (2x-1)^n \exp \left[ - \frac{m_1^2x + m_2^2(1-x)}{M^2 x(1-x)} \right] \left\{ \frac{n+1}{M^2} \left( \frac{1}{12\pi} n(n-1) (2x-1)^{-2} x(1-x) \right.\right. \nonumber\\
&+& \left. \frac{1}{8\pi} \right) + \frac{1}{M^4 x^2(1-x)^2} \left( -\frac{1}{24\pi} \left[ nm_1^2x^3 + nm_2^2(1-x)^3 + 2m_1m_2(nx(1-x)-x^2-(1-x)^2) \right] \right. \nonumber\\
&+& \frac{1}{12\pi} n(n-1) (2x-1)^{-2} x^2(1-x)^2 (m_1+m_2) [m_1x+m_2(1-x)] + \frac{1}{8\pi} x(1-x) \left[ m_1^2x + m_2^2(1-x) \right. \nonumber\\
&+& \left.\left. 2m_1m_2 \right] \right) + \frac{1}{2M^6 x^3(1-x)^3} \left( -\frac{1}{12\pi} \left[ m_1^4x^4 + m_2^4(1-x)^4 + m_1m_2x(1-x)(m_1^2x+m_2^2(1-x)) \right.\right. \nonumber\\
&+& \left.\left.\left. m_1^2m_2^2 x(1-x) (x^2+(1-x)^2) \right] \right) \right\}, \nonumber\\
I^{\rm PS}_{\rm HP,\left<G^3\right>}(M^2) &=& \left<g_s^3fG^3\right> \int^1_0 (2x-1)^n \exp \left[ - \frac{m_1^2x + m_2^2(1-x)}{M^2 x(1-x)} \right] \left\{ \frac{1}{M^4 x^2(1-x)^2} \left( \frac{-9n}{1024\pi^2} \left[ \frac{1}{32}n(n-1)(2x-1)^{-2} \right.\right.\right. \nonumber\\
&+& \left. 6 -\frac{1}{32}n(n-1) + x(1-x) \left( -\frac{1}{8}n(n+63) - 22 + \frac{3}{2} (n+1)(n+6)x(1-x) \right) \right] + \frac{7}{2304\pi^2}n(n-1) \nonumber\\
&\times& (2x-1)^{-2} x(1-x) \left[ 2(n+2)x(1-x) - 3 \right] + \frac{1}{144\pi^2} n(n-1) (2x-1)^{-2} x(1-x) [x^2+(1-x)^2] \nonumber\\
&-& \frac{1}{576\pi^2} (n+1)n^2(n-1) (2x-1)^{-2} x^3(1-x)^3 - \frac{7}{1536\pi^2} n x(1-x) [3 - 2(n+1)x(1-x)] - \frac{1}{96\pi^2} n \nonumber\\
&\times& (n-1) (2x-1)^{-2} x^2(1-x)^2 - \frac{1}{144\pi^2} x(1-x) \left[ -\frac{1}{4} n(n-1) (2x-1)^{-2} + \frac{1}{4}n(n-1) - 3 - 6n \left(n \right.\right. \nonumber\\
&+& \left.\left.\left. 1\right) x(1-x) \right] \right) + \frac{1}{2M^6 x^3(1-x)^3} \left( \frac{-1}{15360\pi^2} \left[ 6m_1^2x \left( \left( \frac{143}{32}n(n-1) - \frac{49}{8}n(n-1)x \right)(2x-1)^{-2} \right.\right.\right. \nonumber\\
&+& \frac{49}{8}(n-1)(n+24)x - \frac{143}{32}n(n-1) + 270 + x(1-x) \left( -\frac{147}{2}(n-1)(n+2)x - \frac{n}{8} (143n+5353) \right. \nonumber\\
&-& \left.\left. 663 + \frac{n}{2}(413n+1477)x(1-x) \right) \right) + 6m_2^2(1-x) \left( \left( \frac{143}{32}n(n-1) - \frac{49}{8}n(n-1)(1-x) \right)(2x-1)^{-2} \right. \nonumber\\
&+& \frac{49}{8}(n-1)(n+24)(1-x) - \frac{143}{32}n(n-1) + 270 + x(1-x) \left( -\frac{147}{2}(n-1)(n+2)(1-x) - \frac{n}{8} \left( 143n \right.\right. \nonumber\\
&+& \left.\left.\left. 5353 \right) - 663 + \frac{n}{2}(413n+1477)x(1-x) \right) \right) + 268m_1m_2 \left( \frac{1}{32}n(n-1)(2x-1)^{-2} - \frac{1}{32}n(n-1) + 3 \right. \nonumber\\
&+& \left.\left. x(1-x) \left( -\frac{1}{8}n(n+47) - 6 + \frac{3}{2}(n-1)(n+4) x(1-x) \right) \right) \right] - \frac{1}{2304\pi^2}n(n-1) (2x-1)^{-2} x(1-x) \nonumber\\
&\times& \left[ 2m_1^2x^2(36x-7) + 2m_2^2(1-x)^2(36(1-x)-7) \right] + \frac{1}{72\pi^2} n(n-1) (2x-1)^{-2} x(1-x) \left[ m_1^2x^3 + m_2^2 \right. \nonumber\\
&\times& \left. (1-x)^3 - m_1m_2x(1-x) \right] - \frac{1}{288\pi^2} n(n-1) (2x-1)^{-2} x^3(1-x)^3 \left[ 2nm_1^2x + 2nm_2^2(1-x) + \left( n \right.\right. \nonumber\\
&-& \left.\left. 1 \right)m_1m_2 \right] - \frac{1}{768\pi^2} x(1-x) \left[ m_1^2x \left( -28nx(1-x) + 37(n-1)x + 21 \right) + m_2^2(1-x) \left( -28nx(1-x) \right.\right. \nonumber\\
&+& \left.\left. 37(n-1)(1-x) + 21 \right) - 32m_1m_2 \left( 1 - (n-1)x(1-x) \right) \right] - \frac{1}{48\pi^2} x(1-x) \left[ m_1^2x \left( -\frac{n}{2}(2x-1)^{-1} \right.\right. \nonumber\\
&-& \left. 8nx(1-x) + (n-1)x - \frac{n+2}{2} \right) + m_2^2(1-x) \left( \frac{n}{2}(2x-1)^{-1} - 8nx(1-x) + (n-1)(1-x) \right. \nonumber\\
&-& \left.\left.\left. \frac{n+2}{2} \right) - 8(n-1) m_1m_2 x(1-x) \right] \right) + \frac{1}{6M^8 x^4(1-x)^4} \left( \frac{-1}{5120\pi^2} \left[ m_1^4x^3 \left( -\frac{147n}{4}(2x-1)^{-1} \right.\right.\right. \nonumber\\
&+& \left. \left( \frac{433n}{2}+1676 \right)x - \frac{147n}{4} - 1080 + x(1-x) \left( -(1263n+1572)x + 385n + 1564 \right) \right) + m_2^4(1-x)^3 \nonumber\\
&\times& \left( \frac{147n}{4}(2x-1)^{-1} + \left( \frac{433n}{2}+1676 \right)(1-x) - \frac{147n}{4} - 1080 + x(1-x) \left( -(1263n+1572)(1-x) \right.\right. \nonumber\\
&+& \left.\left. 385n + 1564 \right) \right) + 4m_1^3m_2x^2 \left( -\frac{87n}{8}(2x-1)^{-1} + \left( \frac{87n}{4} + 375 \right)x - \frac{87n}{8} - 201 + x(1-x) \right. \nonumber\\
&\times& \left. \left( -(221n+161)x + \frac{181n}{2} + 161 \right) \right) + 4m_1m_2^3(1-x)^2 \left( \frac{87n}{8}(2x-1)^{-1} + \left( \frac{87n}{4} + 375 \right) (1-x) \right. \nonumber\\
&-& \left. \frac{87n}{8} - 201 + x(1-x) \left( -(221n+161)(1-x) + \frac{181n}{2} + 161 \right) \right) + 6m_1^2m_2^2 x(1-x) \left( (421n+524) \right. \nonumber\\
&\times& \left.\left. x^2(1-x)^2 - (147n+703)x(1-x) + 147 \right) \right] - \frac{1}{144\pi^2} n(n-1) (2x-1)^{-2} x^3(1-x)^3 \left[ (m_1+m_2) \right. \nonumber\\
&\times& \left. (m_1x+m_2(1-x)) (m_1^2x + m_2^2(1-x)) \right] - \frac{1}{256\pi^2} x(1-x) \left[ m_1^4x^3 (14x+23) + m_2^4(1-x)^3 \left( 14(1-x) \right.\right. \nonumber\\
&+& \left. 23 \right) + 2m_1^3m_2x^2 (-16x+45) + 2m_1m_2^3(1-x)^2 (-16(1-x)+45) + m_1^2m_2^2 x(1-x) \left( -28x(1-x) \right. \nonumber\\
&+& \left.\left. 37 \right) \right] - \frac{1}{16\pi^2} x(1-x) \left[ m_1^4x^3(4x-3) + m_2^4(1-x)^3(4(1-x)-3) - 8m_1^3m_2x^2(1-x) - 8m_1m_2^3x \right. \nonumber\\
&\times& \left.\left. (1-x)^2 - m_1^2m_2^2 x(1-x) (1+8x(1-x)) \right] \right) + \frac{1}{1280\pi^2 \times 24M^{10} x^5(1-x)^5} \left[ m_1^6x^5 \left( 429x(1-x) \right.\right. \nonumber\\
&-& \left. 8x - 135 \right) + m_2^6(1-x)^5 \left( 429x(1-x) - 8(1-x) - 135 \right) + 2m_1^5m_2x^4 \left( 241x(1-x) - 32x - 67 \right) \nonumber\\
&+& 2m_1m_2^5(1-x)^4 \left( 241x(1-x) - 32(1-x) - 67 \right) + m_1^4m_2^2x^3(1-x) \left( 1287x^2(1-x) - 850x(1-x) \right. \nonumber\\
&-& \left. 437x + 294 \right) + m_1^2m_2^4x(1-x)^3 \left( 1287x(1-x)^2 - 850x(1-x) - 437(1-x) + 294 \right) + 4m_1^3m_2^3x^2 \nonumber\\
&\times& \left.\left. (1-x)^2 (-241x(1-x) + 87) \right] \right\}.
\nonumber\\
{\rm Im} I^{\rm PT}_{\rm HP,pert}(s) &=& \frac{-3}{16\pi (n+1)(n+2)(n+3)} (n+1) \left\{ \left( \frac{-m_1^2 + m_2^2 + \bar{s}v}{s} \right)^{n+2} \left[ -m_1^2 + m_2^2 - (n+2) \bar{s}v \right] \right. \nonumber\\
&-& \left. \left( \frac{-m_1^2 + m_2^2 - \bar{s}v}{s} \right)^{n+2} \left[ -m_1^2 + m_2^2 + (n+2) \bar{s}v \right] \right\}, \nonumber\\
I^{\rm PT}_{\rm HP,\left<G^2\right>}(M^2) &=& -\left<\alpha_sG^2\right> (n+1) \int^1_0 dx (2x-1)^n \exp \left[ -\frac{m_1^2x + m_2^2(1-x)}{M^2x(1-x)} \right] \left\{ \frac{1}{M^2} \left(- \frac{1}{12\pi} n(n-1) (2x-1)^{-2} x \right.\right. \nonumber\\
&\times& \left.\left. (1-x) - \frac{1}{24\pi} \right) + \frac{1}{M^4 x^2(1-x)^2} \left( \frac{1}{24\pi} \left[ m_1^2x^3 + m_2^2(1-x)^3 \right] - \frac{1}{12\pi} m_1m_2 x(1-x) \right) \right\}, \nonumber\\
I^{\rm PT}_{\rm HP,\left<G^3\right>}(M^2) &=& -\left<g_s^3fG^3\right> (n+1) \int^1_0 dx (2x-1)^n \exp \left[ -\frac{m_1^2x + m_2^2(1-x)}{M^2x(1-x)} \right] \left\{ \frac{1}{M^4 x^2(1-x)^2} \left( \frac{9}{1024\pi^2} \left[ \frac{1}{32} n(n-1) \right.\right.\right. \nonumber\\
&\times& \left. (2x-1)^{-2} - \frac{1}{32} n(n-1) + 6 + x(1-x) \left( -\frac{1}{8} n(n+63) - 22 + \frac{3}{2} (n+1)(n+6) x(1-x) \right) \right] \nonumber\\
&+& \frac{7}{2304\pi^2} (n+1) x(1-x) [3 - 2(n+4)x(1-x)] + \frac{1}{144\pi^2} x(1-x) \left[ 4(n+1) x(1-x) - n - 2 \right] \nonumber\\
&+& \frac{1}{576\pi^2} (n+1)n(n-1) (2x-1)^{-2} x^3(1-x)^3 + \frac{7}{4608\pi^2} x(1-x) [2(n+1)x(1-x)-3] + \frac{1}{144\pi^2} \nonumber\\
&\times& \left. (n+1) x^2(1-x)^2 \right) + \frac{1}{2M^6 x^3(1-x)^3} \left( \frac{1}{2560\pi^2} \left[ m_1^2x^2 \left( -\frac{49n}{8} (2x-1)^{-1} - \frac{49n}{8} - 180 + \left( \frac{49n}{4} \right.\right.\right.\right. \nonumber\\
&+& \left.\left. 327 \right)x + x(1-x) \left( -(139n+405)x + \frac{131n}{2} + 258 \right) \right) + m_2^2(1-x)^2 \left( \frac{49n}{8} (2x-1)^{-1} - \frac{49n}{8} \right. \nonumber\\
&-& \left.\left. 180 + \left( \frac{49n}{4} + 327 \right)(1-x) + x(1-x) \left( -(139n+405)(1-x) + \frac{131n}{2} + 258 \right) \right) \right] + \frac{1}{1152\pi^2} \nonumber\\
&\times& x(1-x) \left[ m_1^2x \left( -(50n+56) x(1-x) + 29nx + 21 \right) + m_2^2(1-x) \left( -(50n+56) x(1-x) + 29n \right.\right. \nonumber\\
&\times& \left.\left. (1-x) + 21 \right) + 2m_1m_2 \left( (15n+16) x(1-x) - 8 \right) \right] + \frac{1}{72\pi^2} x(1-x) \left[ m_1^2x \left( \frac{n}{4}(2x-1)^{-1} + (2n+4) \right.\right. \nonumber\\
&\times& \left. x(1-x) - \frac{3n}{2}x + \frac{n}{4} - 2 \right) + m_2^2(1-x) \left( -\frac{n}{4}(2x-1)^{-1} + (2n+4)x(1-x) - \frac{3n}{2}(1-x) + \frac{n}{4} - 2 \right) \nonumber\\
&-& \left. 9nm_1m_2x(1-x) \right] + \frac{11}{192\pi^2} n m_1m_2 x^2(1-x)^2 + \frac{1}{288\pi^2} n(n-1) (2x-1)^{-2} x^3(1-x)^3 \left[ m_1^2x + m_2^2 \right. \nonumber\\
&\times& \left. (1-x) \right] - \frac{7}{2304\pi^2} x(1-x) \left[ m_1^2x^2 (2x+1) + m_2^2(1-x)^2 (2(1-x) + 1) \right] - \frac{1}{24\pi^2} m_1m_2 x^2(1-x)^2 \nonumber\\
&+& \left. \frac{1}{72\pi^2} x^2(1-x)^2 \left[ m_1^2x + m_2^2(1-x) + 3m_1m_2 \right] \right) + \frac{1}{6M^8 x^4(1-x)^4} \left( \frac{1}{5120\pi^2} \left[ m_1^4x^4 \left( -429x(1-x) \right.\right.\right. \nonumber\\
&+& \left.\left. 8x + 135 \right) + m_2^4(1-x)^4 \left( -429x(1-x) + 8(1-x) + 135 \right) + 6m_1^2m_2^2 x^2(1-x)^2 \left( 143x(1-x) - 49 \right) \right] \nonumber\\
&+& \frac{1}{384\pi^2} x(1-x) \left[ m_1^4x^3 (36x-7) + m_2^4(1-x)^3 (36(1-x)-7) + m_1^3m_2x^2 (14x+15) + m_1m_2^3(1-x)^2 \right. \nonumber\\
&\times& \left. (14(1-x)+15) + m_1^2m_2^2 x(1-x) (29 - 72x(1-x)) \right] - \frac{1}{12\pi^2} x(1-x) (m_1^2x + m_2^2(1-x)) \left[ m_1^2x^3 \right. \nonumber\\
&+& \left.\left.\left. m_2^2(1-x)^3 + 3m_1m_2x(1-x) \right] + \frac{11}{64\pi^2} x^2(1-x)^2 m_1m_2 [m_1^2x + m_2^2(1-x)] \right) \right\}.
\nonumber
\end{eqnarray}
\end{widetext}
Where $\bar{s} = s - (m_1-m_2)^2$ and $v^2 = 1 - 4m_1m_2/\bar{s}$.

\end{document}